\documentclass{agujournal}

\usepackage{amsmath,amssymb,amsthm}
\usepackage{caption}
\usepackage{comment}
\usepackage{placeins}

\journalname{Space Weather}

\begin{document}

%
%


\title{A variational approach to Data Assimilation in the Solar Wind}

%
%




\authors{Matthew Lang \affil{1}, Mathew J. Owens \affil{2}}


\affiliation{1}{Le Laboratoire des Sciences du Climat et de l'Environnement, CEA-CNRS-UVSQ, 91191 Gif Sur Yvette, France}
\affiliation{2}{Department of Meteorology, University of Reading, Reading, Berkshire, UK}






\correspondingauthor{Matthew Lang}{matthew.lang@lsce.ipsl.fr}




\begin{keypoints}
\item This paper develops a variational data assimilation (DA) method and applies it to a simple solar wind propagation model.
\item Such a DA scheme enables the inner boundary of the solar wind model to be updated using observations in near-Earth space.
\item Experiments performed with both synthetic and STEREO observations show that the DA method is able to reduce errors in the solar wind speeds.
\end{keypoints}

%
%


\begin{abstract}
Variational Data Assimilation (DA) has enabled huge improvements in the skill of operational weather forecasting. In this study, we use a simple solar-wind propagation model to develop the first solar-wind variational DA scheme. This scheme enables solar-wind observations far from the Sun, such as at 1 AU, to update and improve the inner boundary conditions of the solar wind model (at $30$ solar radii). In this way, observational information can be used to improve estimates of the near-Earth solar wind, even when the observations are not directly downstream of the Earth. Using controlled experiments with synthetic observations we demonstrate this method's potential to improve solar wind forecasts, though the best results are achieved in conjunction with accurate initial estimates of the solar wind. The variational DA scheme is also applied to STEREO in-situ observations using initial solar wind conditions supplied by a coronal model of the observed photospheric magnetic field. We consider the period Oct 2010-Oct 2011, when the STEREO spacecraft were approximately $80^{\circ}$ ahead/behind Earth in its orbit. For 12 of 13 Carrington Rotations, assimilation of STEREO data improves the near-Earth solar wind estimate over the non-assimilated state, with a $18.4\%$ reduction in the root-mean-squared-error. The largest gains are made by the DA during times when the steady-state assumption of the coronal models breaks down. While applying this pure variational approach to complex solar-wind models is technically challenging, we discuss hybrid DA approaches which are simpler to implement and may retain many of the advantages demonstrated here.
\end{abstract}

\section{Introduction}

\FloatBarrier
\begin{center}
	\begin{figure}
		\begin{center}
			\includegraphics[width=\textwidth]{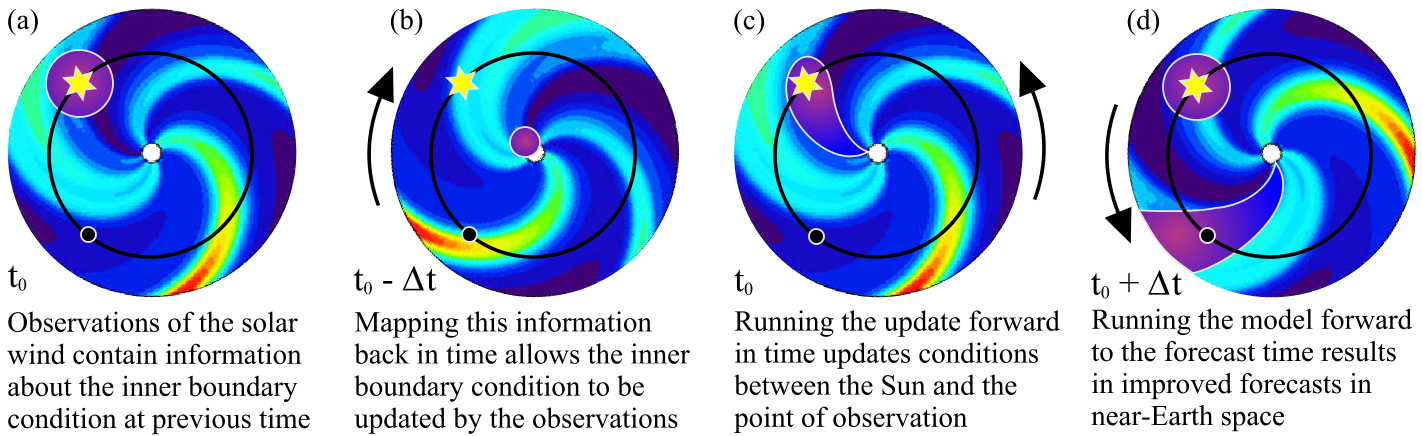}
			\caption{A schematic of a DA scheme that updates the model inner boundary (the white circle) on the basis of observations from a position behind Earth in its orbit (the yellow star). This enables the updated model conditions (the purple regions) to persist until solar rotation brings them to the forecast point at Earth's location (the black circle). Without this ability, any localised change to the model state is quickly lost from the model domain by the super-sonic radial solar wind flow.}
			\label{fig:smootherUpdate}
		\end{center}
	\end{figure}
\end{center}

In meteorology, data assimilation has long been used to improve initial conditions for forecasting, leading to a reduction in the `butterfly effect' and hence improvements in forecasting skill. Furthermore, improvements in the implementation of data assimilation methods into numerical weather prediction models have led to huge improvements in the forecasting accuracy of longer lead-times over the past $20-30$ years \citep{kalnay2003atmospheric}. However, space weather forecasting has yet to exploit the great potential available from implementing data assimilation methods into their forecasting models.

For space-weather forecasting, data assimilation has been attempted in three main domains: the photosphere, the solar wind and the ionosphere. Ionospheric data assimilation is arguably the most mature and as the number of observations of the ionosphere increases, so does its importance and effectiveness. Various data assimilation methods have been applied to the ionosphere, such as 3DVar \citep{bust2008Ion}, 4DVar \citep{wang2004Iono} and Local Ensemble Transform Kalman Filter (LETKF) \citep{durazo2017local}. Photospheric data assimilation, such as the Air Force Data Assimilative Photospheric Flux Transport (ADAPT) model \citep{arge2010air}, uses observations of the magnetic field at the Sun's surface with physics-based temporal evolution to improve the inner boundary condition to coronal models. The improved representation of the corona can in turn be used to generate improved inner-boundary conditions for solar wind models.

In this study, we are looking to exploit observations of the solar wind itself to further improve the inner-boundary conditions for solar wind models (and, as a by product, provide an observationally constrained validation dataset for the outer boundary of coronal models). Previous studies applying data assimilation to in-situ observations of the solar wind, such as \citet{lang2017SWDA}, have focussed on using ensemble-based data assimilation methods to improve the representation of the modelled solar wind. The advantage of these methods is that they are relatively easy to implement for complex numerical models and recent developments have allowed these methods to be incorporated via Message Passing Interface, MPI \citep{browne2015simple, nerger2005pdaf}, in parallel to the numerical models. However, as shown in \citet{lang2017SWDA}, in order to improve the forecast of solar wind conditions in near-Earth space, these methods require observations downstream of the Earth, which are not routinely available. This is due to the localisation of the data assimilation, which means any improvements to the model state from the observations are swept out of the model domain due to the continual radial outflow of the solar wind. In order to make a persistent change to the model state, the model inner boundary conditions must be updated. This is shown schematically in Figure \ref{fig:smootherUpdate}. Ensemble-based Kalman filter techniques cannot propagate information back in time (i.e., back towards the Sun), meaning they cannot be used in this way.

This study investigates the possibility of using variational data assimilation (DA) methods (e.g., \citet{telegrand1986,courtier1994strategy}) for assimilating in-situ observations of the solar wind, specifically to update the inner boundary conditions of the solar wind model. This is achieved using an adjoint method \citep{errico1997adjoint} to map information from the point of observation back to the model's inner boundary. Such changes to the model state would be persistent and remain within the model domain, long after the observation's timestep. A simple solar wind model is described in the next section, followed by a description, in general, of variational data assimilation and a derivation of the particular scheme used in this study. Numerical twin experiments are presented using synthetic observations to test if the method can reconstruct solar wind speed structures and time-series over one solar rotation (i.e. one Carrington Rotation). The data assimilation method is then applied to real spacecraft data, assimilating Solar-Terrestrial Relations Observatory (STEREO) \citep{kaiser2008stereo} solar wind observations from a time when the spacecraft were well separated from Earth (2010-2011), and results verified against Advanced Composition Explorer (ACE) \citep{stone1998advanced} observations in near-Earth space.

Future developments are discussed in the concluding section. Here we note, however, that while the requirement for an adjoint model means that the variational approach is not expected to be practical for a full magnetohydrodynamic (MHD) solar wind model like Enlil \citep{odstrcil2003modeling}, it is nevertheless of interest for two reasons. Firstly, it is useful in developing new solar wind data assimilation techniques with simpler models, as described in Section \ref{sec:expsetup}, which may be valuable in forecast situations. Secondly, it enables us to test whether updating the inner-boundary conditions of the solar wind model is indeed effective in improving model forecast capability, as demonstrated in Section \ref{sec:stereo}. It is hoped that in the future, ensemble-based Kalman smoother methods may be able to provide many of the benefits of variational DA without the need for an adjoint, and thus can be readily applied to MHD models of the solar wind.

\section{Solar wind}

The solar wind is a continuous outflow of plasma and magnetic flux which fills the heliosphere (e.g., \citet{owens2013heliospheric}). The solar wind becomes super-magnetosonic within $10-20$ solar radii ($r_S=695,508 km$). Thus forecasting the near-Earth solar wind conditions is normally treated as a boundary-value problem, with the near-Sun conditions fixed using empirical relations to the coronal magnetic field \citep{mcgregor2011, riley2015}, which itself is determined by extrapolation from the observed photospheric magnetic field (e.g., \citet{mackay2012,linker1999}). The solar wind is then propagated to Earth, typically by a numerical magnetohydrodynamic (MHD) model such as Enlil \citep{odstrcil2003modeling}, with no further observational constraints. In situ spacecraft provide single-point measurements of the solar wind and heliospheric magnetic field which can potentially be used to constrain the solar wind model.

In principle, the DA framework developed here will also be applicable to other solar wind observations, such as Interplanetary Scintillation (IPS; e.g., \citet{breen2006}) or Heliospheric Imagers (HI; e.g., \citet{eyles2009}). These remote measurements of solar wind density structures are subject to line-of-sight integration effects, and estimation of solar wind speed further requires some form of correlation tracking. Thus relative to (single-point) in-situ observations, there is increased uncertainty in both the measurement and its location, but with the advantage of a more synoptic picture of the solar wind. By explicitly accounting for observational errors, a solar wind DA scheme can exploit the positives of both forms of data. Though determining the observational errors is a significant task which is not addressed in the current study.

In the next section we describe a simple solar wind propagation tool which permits more rapid development of DA techniques than the complex and computationally expensive full-MHD approaches \citep{odstrcil2003modeling}. This approach was recently used to explore the forecast potential of large ensembles of solar wind solutions with perturbed initial conditions \citet{owens2017probabilistic}, as is routinely used for operational Numerical Weather Prediction (NWP).

\section{Solar wind propagation model}
\label{sec:solarModel}

In this study, we use the solar wind propagation model of \citet{riley2011mapping}, which maps the equatorial (i.e., two dimensional) solar wind speed over the heliocentric domain from $30 r_S$ to $215 r_S$ from the Sun:
\begin{equation}
\label{eq:modelEvEq1}
v_{i+1,j}(\phi)=v_{i,j}+\frac{\Delta r \Omega_{ROT}}{v_{i,j}}\left(\frac{v_{i,j+1}-v_{i,j}}{C\Delta \phi}\right)
\end{equation}
where $v_{i,j}$ is the speed (in $km/s$) at radius, $r_i$ (the $i$ is the radius coordinate), and at Carrington longitude, $\phi_j$ (where $j$ is the longitude coordinate). Using the same setup as \citet{owens2017probabilistic}, $\Delta r=1 r_S$ is the radial grid resolution (in km), $\Delta \phi=2.81^{\circ}$ is the latitudinal grid resolution, $C=\frac{2 \pi}{180}$ is a constant representing the conversion factor from degrees to radians and $\Omega_{ROT}=\frac{2\pi}{25.38(86400)} s^{-1}$ is the solar rotational speed.

After this solution is obtained, an additional term, $v^{acc}_{i,j}$, is added to the $v_{i,j}$ to represent the acceleration of the solar wind in the domain considered, which is given by:
\begin{equation}
\label{eq:vAccTerm1}
v^{acc}_{i,j}=\alpha v_{0,j}(\phi) \left(1-e^{\frac{r_i}{r_H}}\right)
\end{equation}
where $v_{0,j}$ is the solar wind speed at the inner-boundary and $\alpha=0.15$ and $r_H=50 r_S$ are constants determined by \citet{riley2011mapping}.

Given the 2-dimensional nature of the solar wind model, in this study we must assume the ecliptic plane to be equatorial. In reality, the ecliptic is inclined by $7.25^\circ$ to the heliographic equator. We note that a solar wind DA scheme in a full 3-dimensional solar wind model could relax this assumption.

\section{Data assimilation}
\label{sec:DAscheme}

Data assimilation is the study of combining prior knowledge from a model of a system with information contained in actual observations of that system in order to obtain an optimal estimate of the truth, including its uncertainty. DA methods can be used to 1) provide better model initial conditions for forecasting (e.g. \citep{phil2015HadCM3EWPF, dee2011eraInterim, clayton2013operational}); 2) generate optimal evolution trajectories for the system to study important physical processes (e.g. \citep{broquet2011european, macbean2016consistent}); and 3) improve the model physics by studying model-observation misfits \citep{lang2016ParsationEst}.

Variational data assimilation refers to the subset of DA methods used extensively in meteorological applications \citep{sasaki1970numerical} that provide an optimal fit over the whole time window (the period of time over which the data assimilation is applied). Variational DA typically aims to correct the variables under consideration at the initial time by making use of all the data available over the entire time window. Sequential assimilation, on the other hand, provides an optimal fit at the end of the window by considering each observation sequentially each time a new observation becomes available. Variational data assimilation methods also tend to find the maximum of the posterior probability distribution (the probability distribution of the state given the observations, see Appendix \ref{sec:appCostFunc} for more details), whereas sequential methods typically seek the mean of the posterior probability distribution. For linear systems, this means that the sequential and variational approaches will lead to the same results at the end of the assimilation window. For non-linear systems, however, the posterior probability distribution may have multiple modes, which may lead to the variational approach getting stuck in a local maxima, as opposed to the global maxima of the system, leading to a poorer analysis. Conversely, multi-modal posterior distributions may lead to the mean of the system occurring in an area of low probability in the posterior distribution, leading to the sequential assimilation approach being sub-standard. In these cases, it is unclear what the `optimal' estimate should be, hence there has been substantial efforts to develop Particle Filters \citep{ades2013exploration,phil2015HadCM3EWPF,zhu2016implicit,PJ2014particle}, which aim to estimate the full posterior probability distribution as opposed to a single `optimal' estimate.

With the development of the 4DVar and adjoint model systems \citep{telegrand1986,lorenc1986analysis}, variational methodologies have become much more viable and efficient methods than optimal interpolation methods based on finite-difference methods to calculate the gradient of the cost function (and the Kalman filter methods that preceded them), particularly for applications within high-dimensional meteorological models with large quantities of observations. The 4DVar methodology is typically used to estimate the initial condition of a model given observations over a fixed time-window. However, the purpose of using a variational approach for the solar wind is to map information contained within observations back closer to the Sun, where we can then alter the inner boundary of the model and then re-compute the solar wind speed in the whole domain. Specifically, we wish to estimate the solar wind speed at all Carrington longitudes at the inner boundary ($30 r_S$) using the observations at greater heliocentric distances (i.e., beyond $30 r_S$ from the Sun, typically at Earth orbit, approximately $215 r_S$).
In the remainder of this section, we indicate the general definitions of data assimilation, followed by specific definitions when applied to the solar wind model. Following this, we explain the methodology behind the variational data assimilation proposed in this paper.

In DA, the variables within a numerical model, $\mathbf{x}$, are described by a $N_x$-dimensional state vector, which contains the values of the quantities of interest at all gridpoints, $N$. The variables within a state vector for meteorological applications could include the temperature field over Africa, the mean sea level pressure over the UK, wind speed/direction, etc., depending upon the purpose of the NWP model. The typical dimension of (number of variables within) the state vector in NWP models is of the order $\approx 10^{9}$ \citep{browne2015simple, tremolet2006accounting}.

For this application, the state vector is defined as a vector that contains the solar wind speed at each Carrington longitude, $\phi_j$, in the model domain for a given radius coordinate, $i$, and is written as:
\begin{equation}
\mathbf{v_i}=\left(v_{i,1},v_{i,2},\dots,v_{i,N}\right)
\end{equation}
where $i=0,\dots,185$ correspond to radii coordinates from the Sun as $r_i = 30 r_S, \dots, 215 r_S$ and $N=128$.

For a state $\mathbf{x}$ (e.g., solar wind speed), an estimate of the initial conditions before any data assimilation is referred to as the prior (or background) state, denoted by $\mathbf{x^b}$. The prior state is generated using available prior information about the state. In meteorological applications, this typically comes from a previous forecast. The prior state is assumed to be a random perturbation away from the true state, such that:
\begin{equation}
\label{eq:ch2:backgroundStateDef}
\mathbf{x_0^t}=\mathbf{x^b}+\boldsymbol{\xi_0}
\end{equation}
where $\mathbf{x_0^t}$ is an $N_x$-dimensional discretisation of the true initial state, and $\boldsymbol{\xi_0}$ represents the random error in the prior state.

For our solar wind model, the prior inner boundary condition, $\mathbf{v_{0}^b}$, is an inner boundary condition that we must provide (i.e. from a previous forecast of the inner boundary, such as from a previous coronal solution). The prior inner boundary condition is assumed to be a random perturbation from the true state, such that $\mathbf{v_{0}^b} \sim \mathcal{N}(\mathbf{v_{0}^t},\mathbf{B})$, where $\mathbf{v_{0}^t}$ represents the true speed at the inner boundary and $\mathbf{B}$ represents the prior error covariance matrix (i.e. the covariance matrix of the errors at the inner boundary).

Observations within the time window provide information about the true state of the system. In order to merge the information contained by the observations with the state generated by the numerical model, it is necessary to define a function that maps from the state space to the observation space. This function is called the observation operator and is defined as:
\begin{equation}
\label{eq:ch2:genObservationOpDef}
\mathbf{y}=\mathcal{H}(\mathbf{x})+\boldsymbol{\epsilon}
\end{equation}
where $\mathcal{H}: \mathbb{R}^{N_x} \rightarrow \mathbb{R}^{N_y}$ maps the state into observation space, with $\mathbf{y}$ being an observation with the observation error given by $\boldsymbol{\epsilon}$.


Thus for the solar wind, the $k^{th}$ vector of observations, $\mathbf{y_k}$, at radius $r_k$, are defined as:
\begin{equation}
\label{eq:obsDef}
\mathbf{y_k}=\mathcal{H}_k\left(\mathbf{v_{i_k}}\right)+\boldsymbol{\epsilon_k}
\end{equation}
where $\mathcal{H}_k$ is the observation operator, a function that maps the model solar wind speed vector to the observation space (i.e. what the observation would be for any given $\mathbf{v}$) and $\boldsymbol{\epsilon_k}$ is the random observation error, assumed to be normally distributed, $\boldsymbol{\epsilon_k} \sim \mathcal{N}(\mathbf{0},\mathbf{R_k})$.

The numerical model, which approximates the dynamics of the system, is denoted by:
\begin{equation}
\label{eq:ch2:genModelEq}
\mathbf{x_{i+1}}=f_i\left(\mathbf{x_{i}}\right)+\boldsymbol{\eta_{i}}
\end{equation}
where $\mathbf{x_{i}}$ is the state vector at point $i$, the numerical model is represented by $f_i$ (which here is the solar wind propagation model described in section \ref{sec:solarModel}), and $\boldsymbol{\eta_{i}}$ is an $N_x$-dimensional term representing the model error. The distribution of the model error is almost always unknown and may contain biases towards particular states (i.e. have non-zero mean) or be multi-modal, etc.

We now adopt the Strong Constraint approach and assume that the numerical model is perfect, i.e. contains no model error \citep{lorenzo2007weakStrong, fisher2005equiv}. In practice, this typically produces a poorer result than the weak constraint solution that allows for model error. But the weak constraint problem is much more complex as it is impossible to know precisely where the model is incorrect, due to missing physics, the effects of sub-grid processes etc., and hence very difficult to prescribe an accurate model error covariance matrix. In addition, including model error can lead to coupled equations that can result in different solutions, depending on which order the equations are solved \citep{parEstEvensen1998,lang2016ParsationEst}, leading to additional work being done to decouple them \citep{bennett1992}. Therefore, as this is an introductory study to demonstrate the effectiveness of the variational approach, it is logical to start with the Strong Constraint approximation. The model evolution equation can then be rewritten as:
\begin{align}
\mathbf{v_{i+1}}&=f_i(\mathbf{v_{i}})\\
&=\left(f_{i,1}(\mathbf{v_i}),f_{i,2}(\mathbf{v_i}),\dots,f_{i,N}(\mathbf{v_i})\right)
\end{align}
where
\begin{equation}
f_{i,j}(\mathbf{v_i})=\left\{ \begin{matrix} v_{i,j}+\frac{\Delta r \Omega_{ROT}} {v_{i,j}} \left(\frac{v_{i,j+1}-v_{i,j}}{C \Delta \phi}\right)+\alpha v_{0,j} \left(e^{\frac{r_{i-1}}{r_H}}-e^{\frac{r_i}{r_H}}\right)  \hspace{2cm}& \text{if $r_i \ne 30 r_S$} \\
v_{i,j}+\frac{\Delta r \Omega_{ROT}} {v_{i,j}} \left(\frac{v_{i,j+1}-v_{i,j}}{C \Delta \phi}\right) +\alpha v_{0,j} \left(1-e^{\frac{r_i}{r_H}}\right) \hspace{2cm} & \text{Otherwise} \end{matrix} \right.
\end{equation}
where $r_i$ is the radial distance from the Sun at radius coordinate, $i=0,\dots,185$. The model evolution equation at the inner boundary, at $30 r_S$, adds the acceleration term onto the primary calculation for the speed at the next radial coordinate. However, for each subsequent radial coordinate, the acceleration term from the previous radial point must be removed prior to the addition of the new acceleration to avoid an accumulation of the acceleration terms within the model.

Use of the Strong Constraint \citep{howes2017strong} allows the observation operator to map from the speed at the inner boundary, $\mathbf{v_{0}}$, to the observation location, such that:
\begin{equation}
\label{ref:obsDefInnBoun}
\mathbf{y_k}=\mathcal{H}_k\left[f_{i_k-1}\left(f_{i_k-2}\left(\dots f_{0}\left(\mathbf{v_{0}}\right)\dots\right)\right)\right]+\boldsymbol{\epsilon_k}
\end{equation}
where $k$ denotes the observation number and the $i_k$ are a subset of the radial coordinates, $i$, and are the radial coordinates where the $k^{th}$ observation occurs.

This allows a cost function, $\mathcal{J}$, to be written purely in terms of the inner boundary solar wind speed, such that:
\begin{align}
\label{eq:costFunc2}
\mathcal{J}(\mathbf{v_{0}})&=\frac{1}{2}\left(\mathbf{v_{0}}-\mathbf{v^{b}_{0}}\right)^T\mathbf{B}^{-1}\left(\mathbf{v_{0}}-\mathbf{v^{b}_{0}}\right) \notag \\
&+ \frac{1}{2}\sum_{k=1}^{N_y}\left(\mathbf{y_k} -\mathcal{H}_k\left[f_{i_k-1}\left(f_{i_k-2}\left(\dots f_{0}\left(\mathbf{v_{0}}\right)\dots\right)\right)\right]\right)^T\mathbf{R_k}^{-1}\left(\mathbf{y_k} -\mathcal{H}_k\left[f_{i_k-1}\left(f_{i_k-2}\left(\dots f_{0}\left(\mathbf{v_{0}}\right)\dots\right)\right)\right]\right)
\end{align}
where $\mathbf{v_{0}}$ is the state vector at the inner boundary (at $30 r_S$).

The cost function is the sum of the relative contributions of the errors present within the solar wind system. Its derivation is outlined in Appendix \ref{sec:appCostFunc}. The first term in the cost function represents the prior errors, $\left(\mathbf{v_{0}}-\mathbf{v^{b}_{0}}\right)$, at the inner boundary weighted by the prior error covariance matrix, $\mathbf{B}^{-1}$. The second term represents the sum of the observation errors at each observed radial coordinate, $\left(\mathbf{y_k} -\mathcal{H}_k\left[f_{r_k-1}\left(f_{r_k-2}\left(\dots f_{0}\left(\mathbf{v_{0}}\right)\dots\right)\right)\right]\right)$, and is weighted by the observation error covariance matrix $\mathbf{R}^{-1}$. Therefore, the optimal state, that minimises the errors in the system, is the state that minimises the cost function. The weighting by the inverse of the error covariance matrices means that if there is a high amount of certainty in, for example, the prior state compared to that of the observations, then $\mathbf{B}^{-1}$ will be much larger than $\mathbf{R}^{-1}$, meaning that the optimal state vector will have to move closer to the prior state to minimise the cost function, increasing the dominance of the prior state on the final analysis.

Obtaining the $\mathbf{v_{0}}$ that minimises the cost function is a non-trivial task. This can be done by evaluating the finite differences of $\mathcal{J}$ or by evaluating $\nabla_{\mathbf{v}} \mathcal{J}$ directly \citep{bannister2007Elementary}. However, these methods are impractical for higher dimensional systems. In such situations, a more efficient method of calculating the gradient is to use the adjoint method, a method of sensitivity analysis that efficiently computes the gradient of a function \citep{errico1997adjoint}. For the Strong Constraint approach, this involves using the method of Lagrange multipliers to generate a set of adjoint equations.

Firstly, we rewrite the cost function that we wish to minimise as:
\begin{equation}
\mathcal{J}(\mathbf{v_{0}})=\frac{1}{2}\left(\mathbf{v_{0}}-\mathbf{v^{b}_{0}}\right)\mathbf{B}^{-1}\left(\mathbf{v_{0}}-\mathbf{v^{b}_{0}}\right)^T + \frac{1}{2}\sum_{k=1}^{N_y}\left(\mathbf{y_k} -\mathcal{H}_k(\mathbf{v_{i_k}})\right)\mathbf{R}^{-1}\left(\mathbf{y_k} -\mathcal{H}_k(\mathbf{v_{i_k}})\right)^T
\end{equation}
subject to the (strong) constraint
\begin{equation}
\mathbf{v_{r+1}}=f_r(\mathbf{v_r}).
\end{equation}

To minimise this cost function subject to the constraint, we must minimise the Lagrangian \citep{arfken2011mathematical}, $\mathcal{L}(\mathbf{v_r},\boldsymbol{\lambda_r})$, which is defined by:
\begin{equation}
\mathcal{L}(\mathbf{v_r},\boldsymbol{\lambda_r})=\mathcal{J}(\mathbf{v_{0}})+\sum_{i=0}^{N_r}\boldsymbol{\lambda}^T_{\mathbf{i+1}}(\mathbf{v_{i+1}}-f_r(\mathbf{v_i})).
\end{equation}
where the $\boldsymbol{\lambda_i}$'s are the Lagrange multipliers.

By differentiating with respect to the $v_r$ and the $\lambda_r$ variables, it is possible to obtain a set of adjoint equations, given by:
\begin{align}
\mathbf{v_{r+1}}&=f_r(\mathbf{v_r})\\
\boldsymbol{\lambda_{N_r+1}}&=\mathbf{0} \label{eq:firstAdjEqn}\\
\boldsymbol{\lambda_r}&=\left\{\begin{matrix}\mathbf{F_r}^T\boldsymbol{\lambda_{r+1}}+\mathbf{H_k}^T\mathbf{R_k}^{-1}(\mathbf{y_k}-\mathcal{H}_k(\mathbf{v_{r}})) \hspace{2cm} & \text{if $r=r_k$ is obs. radius} \\
\mathbf{F_r}^T\boldsymbol{\lambda_{r+1}} \hspace{2cm} & \text{Otherwise}
\end{matrix}\right.\\
\boldsymbol{\lambda_0}&=\mathbf{F_{0}}^T\boldsymbol{\lambda_1}+\mathbf{B}^{-1}\left(\mathbf{v_0}-\mathbf{v_0^b}\right)
\end{align}
where $\mathbf{F_r}$ is given by the Jacobian of $f_r$, such that:
\begin{equation}
\mathbf{F_r}=\begin{pmatrix}
\frac{\partial f_{r,1}}{\partial v_{r}(\phi_1)} & \frac{\partial f_{r,1}}{\partial v_{r}(\phi_2)} & \dots & \frac{\partial f_{r,1}}{\partial v_{r}(\phi_N)}\\
\frac{\partial f_{r,2}}{\partial v_{r}(\phi_1)} & \frac{\partial f_{r,2}}{\partial v_{r}(\phi_2)} & \dots & \frac{\partial f_{r,2}}{\partial v_{r}(\phi_N)}\\
\vdots & \vdots & \ddots & \vdots \\
\frac{\partial f_{r,N_r}}{\partial v_{r}(\phi_{N_r})} & \frac{\partial f_{r,N_r}}{\partial v_{r}(\phi_2)} & \dots & \frac{\partial f_{r,N_r}}{\partial v_{r}(\phi_N)}\\
\end{pmatrix}
\end{equation}
and
\begin{equation}
\frac{\partial f_{r,i}}{\partial v_{r}(\phi_j)}=\left\{ \begin{matrix}
1-\frac{\Delta r \Omega_{ROT}}{C \Delta\phi}\left(\frac{v_{r}(\phi_{i+1})}{v^{2}_{r}(\phi_i)}\right) \hspace{2cm} & \text{if $\rho_r>30r_S$ and $j=i$}\\
\frac{\Delta r \Omega_{ROT}}{C \Delta{\phi}}\left(\frac{1}{v_{r}(\phi_i)}\right) \hspace{2cm} & \text{if $\rho_r>30r_S$ and $j=i+1$}\\
1-\frac{\Delta r \Omega_{ROT}}{C \Delta{\phi}}\left(\frac{v_{0}(\phi_{i+1})}{v^2_{0}(\phi_i)}\right)+\alpha\left(1-e^{\frac{\rho_r}{\rho_H}}\right) \hspace{2cm} & \text{if $\rho_r=30r_S$ and $j=i$}\\
\frac{\Delta r \Omega_{ROT}}{C \Delta{\phi}}\left(\frac{1}{v_{r}(\phi_i)}\right) +\alpha\left(1-e^{\frac{\rho_r}{\rho_H}}\right) \hspace{2cm} & \text{if $\rho_r=30r_S$ and $j=i+1$}
\end{matrix}\right.
\end{equation}

The Lagrange multiplier at the inner boundary radius gives us the negative of the gradient at $\mathbf{v_{0}}$, the value we wish to find. This implies that:
\begin{equation}
\nabla_{\mathbf{v_{0}}} \mathcal{J}(\mathbf{v_0})=-\boldsymbol{\lambda_0}=-\mathbf{F_{0}}^T\boldsymbol{\lambda_1}-\mathbf{B}^{-1}\left(\mathbf{v_{0}}-\mathbf{v^b_{0}}\right) \label{eq:lastAdjEqn}
\end{equation}

This gradient allows the use of a plethora of available gradient-based minimisation algorithms, from Newton's methods to steepest descent methods \citep{bazarra2013nonlin}. In this study, however, we use the $\text{Broyden}-\text{Fletcher}-\text{Goldfarb}-\text{Shanno}$ (BFGS) algorithm, a Quasi-Newtonian method, that is commonly used for its speed and accuracy, even for non-linear problems.

\section{Experimental setup}
\label{sec:expsetup}
\begin{figure}
	
	\includegraphics[width=\textwidth]{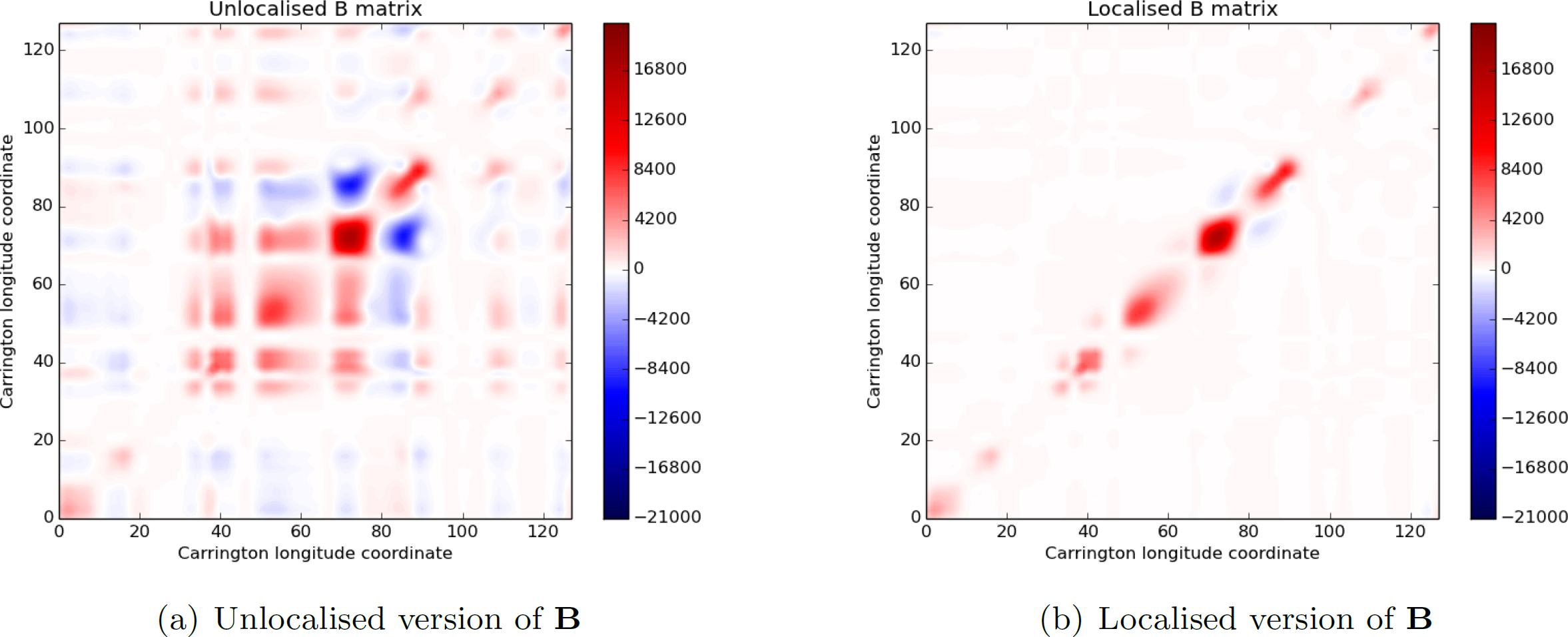}
	\caption{The generation of the prior error covariance matrix from the $576$-member ensemble is shown in $(a)$. The effects of the localisation with a $15^{\circ}$ localisation length scale are shown in $(b)$.}
	\label{fig:Bunloc}
\end{figure}

For this data assimilation method, we must provide a set of conditions in order to proceed. These include specifying the prior error covariance matrix and the observation error covariance matrix. The specification of these conditions is a very important part of the data assimilation process. For example, if the specified prior error uncertainty is too small, the data assimilation analysis will have too much confidence in the quality of the prior state and the data assimilated analysis state will have less freedom to move away from it.

In meteorology, the prior error covariance matrix, $\mathbf{B}$, is built up by operational centres studying the misfits between forecasts and reanalysis datasets. In comparison, the space weather forecasting is a relatively young science and we do not have this back-catalogue of available forecasting and reanalysis datasets to compare against. Therefore, a different interim approach must be taken. (Full specification of $\mathbf{B}$ for the solar wind will be the study of future study and here we make an initial estimate to permit progress at this stage.)

We here approximate the prior error covariance matrix using an ensemble of solar wind speed states generated in the same way as \citet{owens2017probabilistic}. The initial conditions to the solar wind propagation model are near-Sun ($30 r_S$) solar wind speeds, derived from Carrington rotation solutions of the MAS (Magnetohydrodynamics Around a Sphere) coronal model \citet{linker1999} in which the solar wind speed at $30 r_S$ is determined using empirical relations to the coronal magnetic field configuration \citep{riley2012corotating}. These data are available from http://www.predsci.com/mhdweb/home.php. Near-Earth solar wind speed is determined using Equations \ref{eq:modelEvEq1} and \ref{eq:vAccTerm1} to propagate the MAS solar wind speed at $30r_S$ at the sub-Earth point to Earth orbit at approximately $215 r_S$. An ensemble of $576$ members is further generated using perturbed initial conditions. This is achieved by sampling the MAS $30 r_S$ solar wind speed at a range of latitudes about the sub-Earth point (an example is shown in Figure 4 of \citet{owens2017probabilistic}) and independently propagating each perturbed set of initial conditions to $215 r_S$. The resulting ensemble of near-Earth solar wind speeds has been shown to provide an accurate measure of the uncertainty in the unperturbed forecast \citep{owens2017probabilistic}.

Here, we use the $576$-member ensemble created using MAS solutions to Carrington rotation $2100$, which spans early August to early September 2010. This interval contains no interplanetary coronal mass ejections, but does contain both fast and slow solar wind. The ensemble is used to approximate the $\mathbf{B}$ matrix, similar to that of the Ensemble Kalman Filter (EnKF) and methods used in other studies, such as \citet{pereira2006ensemble}. The $\mathbf{B}$ matrix is approximated by:
\begin{equation}
\mathbf{B} \approx \frac{1}{M-1} \sum_{m=1}^{M}\left[\left(\mathbf{v^{(m)}_0}-\overline{\mathbf{v^M_0}}\right)\left(\mathbf{v^{(m)}_0}-\overline{\mathbf{v^M_0}}\right)^T\right]
\end{equation}
where $\overline{\mathbf{v^M_0}}=\frac{1}{M}\sum_{m=1}^{M}\left[v^{(m)}_0\right]$ and $M=576$.

A common issue in generating the error covariance matrix using a (finite) ensemble is that spurious correlations are introduced, as can be seen in Figure \ref{fig:Bunloc}. The effect of this is that an observation may have an unrealistic impact on a model state variable at a distant location and degrade the quality of the analysis. Indeed, \citet{hamill2001distance} showed that if the error in the covariance estimate provided by the ensemble (the noise) is greater than the true correlation (the signal), the accuracy of an EnKF analysis would decrease. Additionally, they show that the signal-to-noise ratio is a function of ensemble size, with larger ensembles representative of the true statistics of the system, with less associated noise. Therefore, it is necessary to minimise the effect of these spurious correlations in the estimated prior error covariance matrix, $\mathbf{B_{ens}}$. To do this, we utilise a technique called localisation, that is used to restrict the covariances to more realistic spatial/temporal-scales and hence reduce the effective noise to signal ratios at longer space/time-scales, increasing the effective size of the ensemble in the process.

As a first approximation, we use a standard distance-based Gaussian localisation scheme (see \citet{lang2017SWDA} for a discussion on possible localisation schemes for solar wind data assimilation), and apply it to the ensemble-estimated covariance matrix, $\mathbf{B_{ens}}$, such that the localised prior error covariance matrix, $\mathbf{B_{loc}}$, becomes:
\begin{equation}
\mathbf{B_{loc}}=\mathbf{L} \circ \mathbf{B_{ens}}
\end{equation}
where $\circ$ represents the Schur-product and $\mathbf{L}$ is the localisation matrix with entries given by
\begin{equation}
L_{i,j}=\exp^{-\frac{(i-j)\Delta \phi}{S}}
\end{equation}
and $S$ is the localisation length scale. As a starting point, we set $S=15^{\circ}$, which is the approximate width of the slow wind band, and hence mimics the large-scale spatial variability in solar wind speed (e.g., \citet{owens2017global}). The choice of this value is somewhat arbitrary and further research is required. However, a preliminary sensitivity analysis suggests that at least within the range $10^\circ$ to $20^\circ$, the results presented below are not significantly affected. A finite $S$ is nevertheless required, as the noise present on the larger spatial-scales ($\ge \approx 60^{\circ}$) can cause the adjoint calculations to become numerically unstable.

The observation error covariance matrix, $\mathbf{R}$, is also an unknown quantity in solar wind modelling. Properly addressing this is a substantial research topic in its own right, far beyond the scope of this initial study. In addition to the measurement uncertainty, $\mathbf{y}$, it also accounts for the uncertainty from the ``representativity" error \citep{janjic2017representation}, resulting from the approximation of a continuous process in discrete space, the error resulting from representing an observation in the incorrect location, and the error resulting from representing a single measurement over the (potentially large) volume of a grid-cell. Additionally, there is also an implicit component of the model error within $\mathcal{H}\left(\mathbf{x}\right)$ which further complicates the specification of this quantity. As a first approximation of the observation error (to be further tested in future work), we begin with the simplest approach, assuming that the observations are all independent of one another (i.e., a diagonal $\mathbf{R}$ matrix). We must also specify an observational error standard deviation, which we set at $10 \%$ of the mean prior solar wind speed at the observation radius. This is the same order as the observed variability of near-Earth solar wind speeds. Again, this is still a somewhat arbitrary value, but it provides a starting point from which we can progress and test such assumptions. Then together, this gives:
\begin{align}
\label{eq:R01xMean}
R_{kk}&=(0.1\overline{\mathbf{v^b_{r_k}}})^{2} \notag\\
&=(0.1 \sum_{i=1}^{N}\left[v^b_{r}(\phi_i)\right])^{2}.
\end{align}

\section{Results}
\subsection{Observing System Simulation Experiments}
\label{sec:OSSEs}

\begin{table}
	\begin{center}
		\begin{tabular}{ | p{2cm} | p{1.5cm} | p{1.5cm} | p{1.2cm} | p{1.5cm} | p{1.5cm} | }
			\hline
			OSSE exp. & Init. $\mathcal{J}(\mathbf{v_0})$ & Final $\mathcal{J}(\mathbf{v_0})$ & No. of iter. & Prior RMSEs (km/s) & Post. RMSEs (km/s) \\ \hline
            ``Optimal" prior, unshifted & 760.879 & 156.364 & 458 & 94.075 & 26.222 \\ \hline
            Prior shifted $62 \Delta \phi$ & 3677.218 & 843.463 & 448 & 217.109 & 87.582 \\ \hline
            Uniform prior & 1661.161 & 684.787 & 456 & 182.486 & 104.320 \\
			\hline
		\end{tabular}
	\end{center}
	\caption{Table showing prior/posterior cost function values, the number of iterations required for convergence and prior/posterior RMSEs for the OSSE experiments, described in section \ref{sec:OSSEs}, using a "good" prior state generated from the same distribution as the true state, a prior state shifted by $62 \Delta \phi \approx 174^{\circ}$ and a uniform prior speed of $500 km/s$.}
	\label{tab:0dPhiPriorTwin}
\end{table}

\begin{figure}
	\begin{center}
		\includegraphics[width=\textwidth]{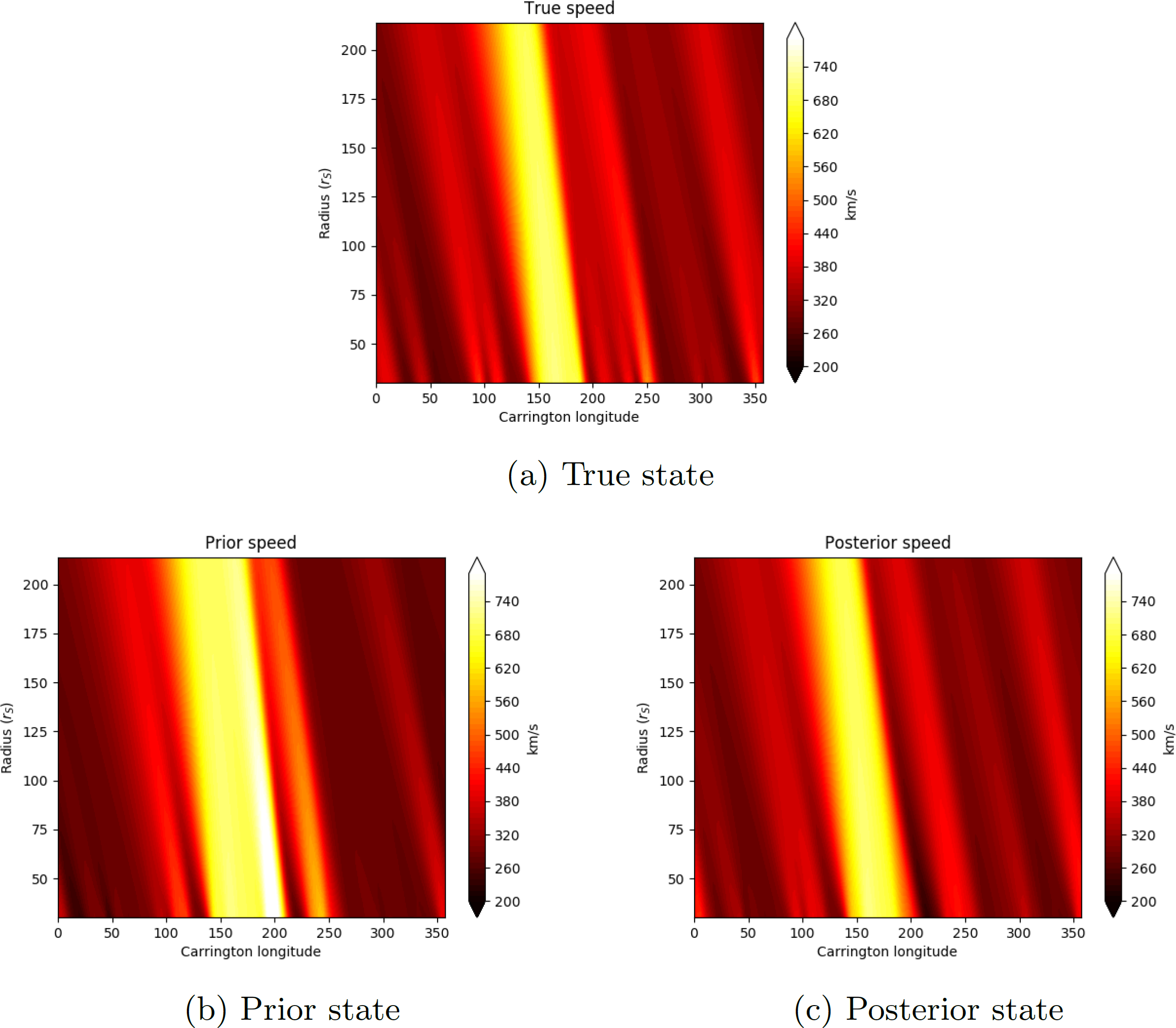}
		\caption{The speed of the solar wind in $km/s$ propagated from the $30r_S$ to the $215r_S$ using the solar wind propagation model. $(a)$ shows the speeds generated from the truth state, $(b)$ shows the speeds generated from the prior state and $(c)$ shows the posterior state after 50 iterations of the forward and adjoint model have been performed.}
		\label{fig:twinPriorState}
	\end{center}
\end{figure}

\begin{figure}
	\begin{center}

		\includegraphics[width=\textwidth]{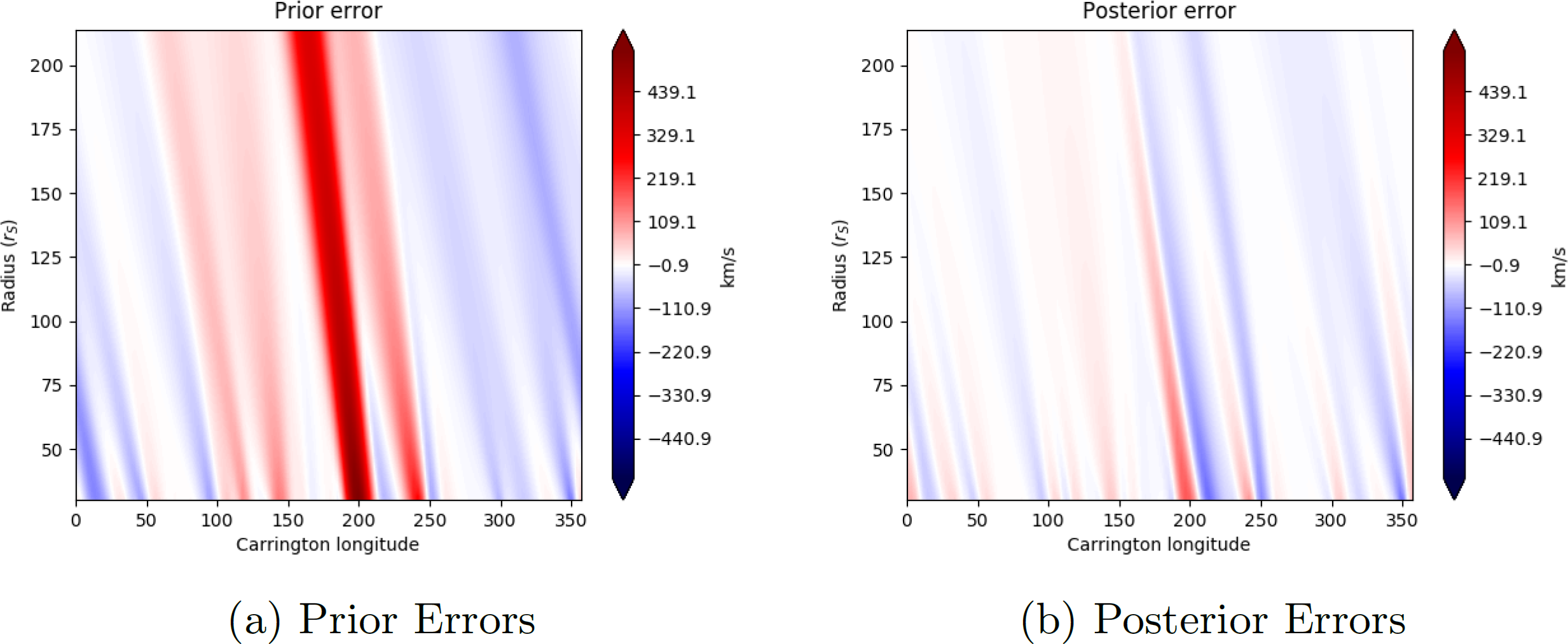}
		\caption{(a) The differences between the prior and the `true' solar wind speed and (b) the differences between the posterior and the true solar wind speed, in $km/s$.}
		\label{fig:twinErrors}
	\end{center}
\end{figure}

This section describes initial tests of the variational data assimilation method derived in the previous sections. In order to perform numerical experiments in a controlled way, we perform observing system simulation experiments (OSSEs, also known as identical twin experiments), a typical verification approach for a data assimilation scheme. An OSSE uses the uncertainty present within the model state to generate two model states that we refer to as the prior state and the \emph{truth} state. Specifically, we draw a prior and \emph{truth} state from the normal distribution, $\mathcal{N}\left(\overline{\mathbf{v^M_0}},\mathbf{B}\right)$.

We use our numerical model to propagate this truth state to get a complete \emph{truth trajectory}. We then take synthetic observations from a fixed point in space from this truth trajectory. These `observations' are perturbed by measurement noise to mimic a real data assimilation experiment. The goal is to generate a posterior state that represents the best estimate of the truth run, using the limited information from the observations and uncertain prior information on the boundary conditions. This posterior state is then compared to the prior state to evaluate the performance of the data assimilation scheme used.


In our experiments, the true state, $\mathbf{v_t}$, is generated as a random sample from the normal distribution of the MAS ensemble for CR2100, $\mathcal{N}\left(\overline{\mathbf{v_0^M}},\mathbf{B_{loc}}\right)$. This true state is propagated out to $215 r_S$ by the solar propagation model detailed in Section \ref{sec:solarModel} and is shown in Figure \ref{fig:twinPriorState}a. The prior state, $\mathbf{v^b_0}$, is also randomly drawn from the same distribution and is shown mapped by the solar wind propagation model to $215 r_S$ in Figure \ref{fig:twinPriorState}b. In this case, the prior state is reasonably close to the true state, with the most obvious difference being the fast stream around 150-200$^\circ$ Carrington longitude is larger in the prior state than in the true state. This is seen as a positive band in the prior error shown in Figure \ref{fig:twinErrors}a.

Direct observations of the true state were taken at $215 r_S$ every $\Delta \phi$, mimicking a time series of near-Earth spacecraft observations every $\sim 5$ hours, which is a lower frequency than that of real observational data. A random perturbation, $\boldsymbol{\epsilon}$, drawn from the distribution $\mathcal{N}(\mathbf{0},\mathbf{R})$ is added to mimic the effects of observation error. The prior state is used as an initial estimate for the `optimal state' that minimises the cost function (equation \eqref{eq:costFunc2}). The gradient of the cost function at this point is calculated by the adjoint equations (equations \eqref{eq:firstAdjEqn}-\eqref{eq:lastAdjEqn}). This gradient is then input into the BFGS minimisation algorithm to obtain an estimate for the minimum of the cost function and the procedure is iterated, using the new estimate as the `optimal state', until the cost function converges to a minimal state, within a tolerance of $10^{-5}$ (i.e. such that the algorithm is repeated as long as the gradient norm of the cost function, from one iteration to the next, is greater than $10^{-5}$).

The posterior state generated by the DA method (i.e., the state that minimises the cost function) is shown in Figure \ref{fig:twinPriorState}c. An improvement over the prior state (in Figure \ref{fig:twinPriorState}b) can be seen in the form of the high speed stream more closely matching that of the true state. Figure \ref{fig:twinErrors} shows that the posterior error does not suffer from the large positive error band of the prior. In addition, there are smaller structures present within the true speed that are recreated in the posterior at around Carrington Longitudes $50^{\circ}, 100^{\circ}, 300^{\circ}$ and $350^{\circ}$. Conversely, the finer structures between $200^{\circ}-250^{\circ}$ cannot be recreated by the DA due to insufficient observational information to overcome the large prior errors in this region.

The root mean-square error for the OSSEs are calculated over the whole domain, calculated as:
\begin{equation}
RMSE_{OSSE}=\frac{1}{N_r}\frac{1}{N}\sum_{i=1}^{N_r}\sum_{j=1}^{N}\left[\left(v_{i,j}-v^t_{i,j}\right)^2\right]
\end{equation}
where $v_{i,j}$ represent the prior or posterior state, dependent upon whether the prior or posterior RMSEs are being generated. Table \ref{tab:0dPhiPriorTwin} shows that there is a reduction of approximately $72\%$ in RMSE as a result of the variational data assimilation analysis.

\subsection{The effect of prior state}

\begin{figure}
\begin{center}
\includegraphics[width=\textwidth]{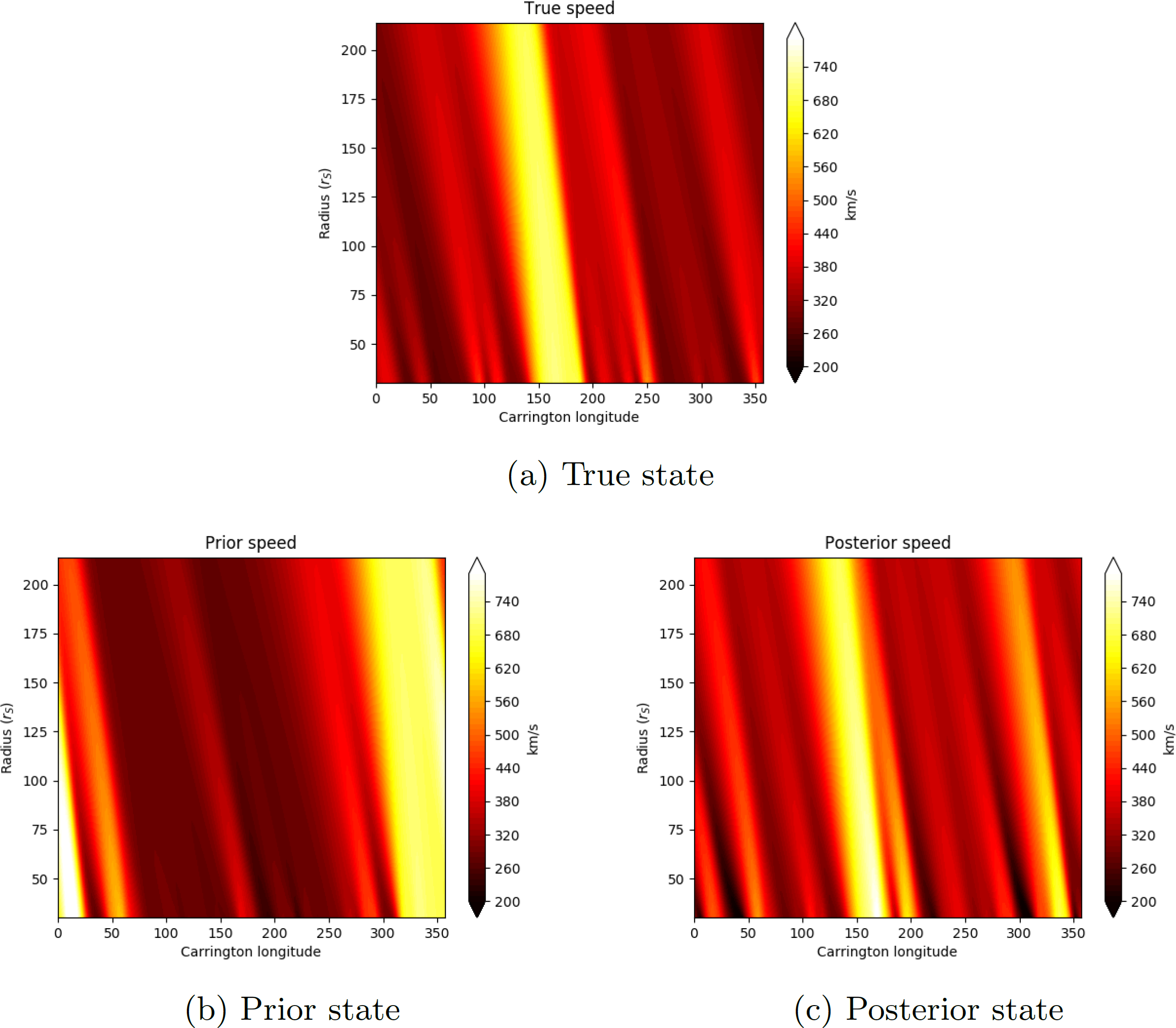}
\caption{Solar wind in $km/s$ propagated from the $30r_S$ to the $215r_S$ using the solar wind propagation model. $(a)$ Speeds generated from the truth state, $(b)$ speeds generated from the prior state that is shifted by $62 \Delta \phi$ and $(c)$ shows the posterior state after the data assimilation has been performed.}
\label{fig:twinPriorState62dPhi}
\end{center}
\end{figure}

\begin{figure}
	\begin{center}
	\includegraphics[width=\textwidth]{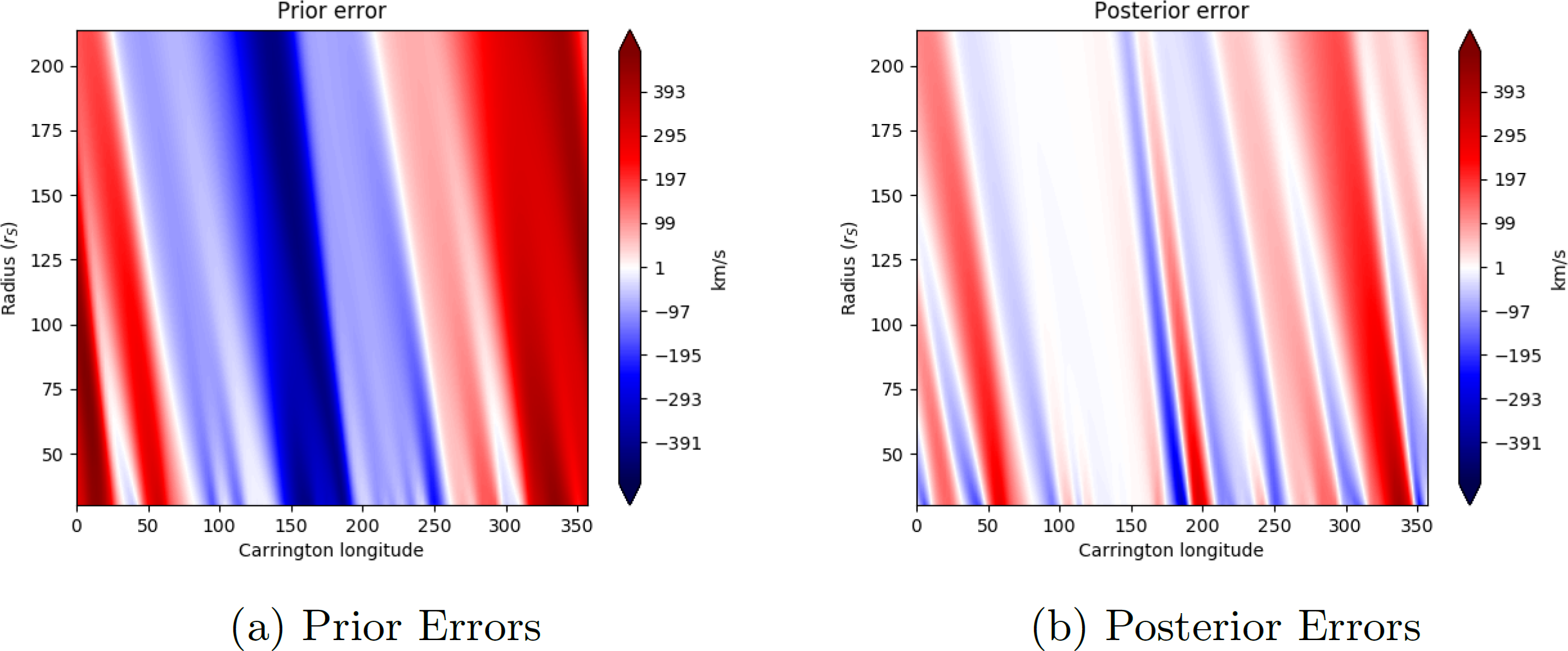}
	\caption{(a) The differences between the prior and the `true' solar wind speeds and $(s)$ the differences between the posterior and the true solar wind speeds, in $km/s$, when the prior state is shifted by $62 \Delta \phi$.}
	\label{fig:twinErrors62dPhi}
\end{center}
\end{figure}

\begin{figure}
\begin{center}
	\includegraphics[width=\textwidth]{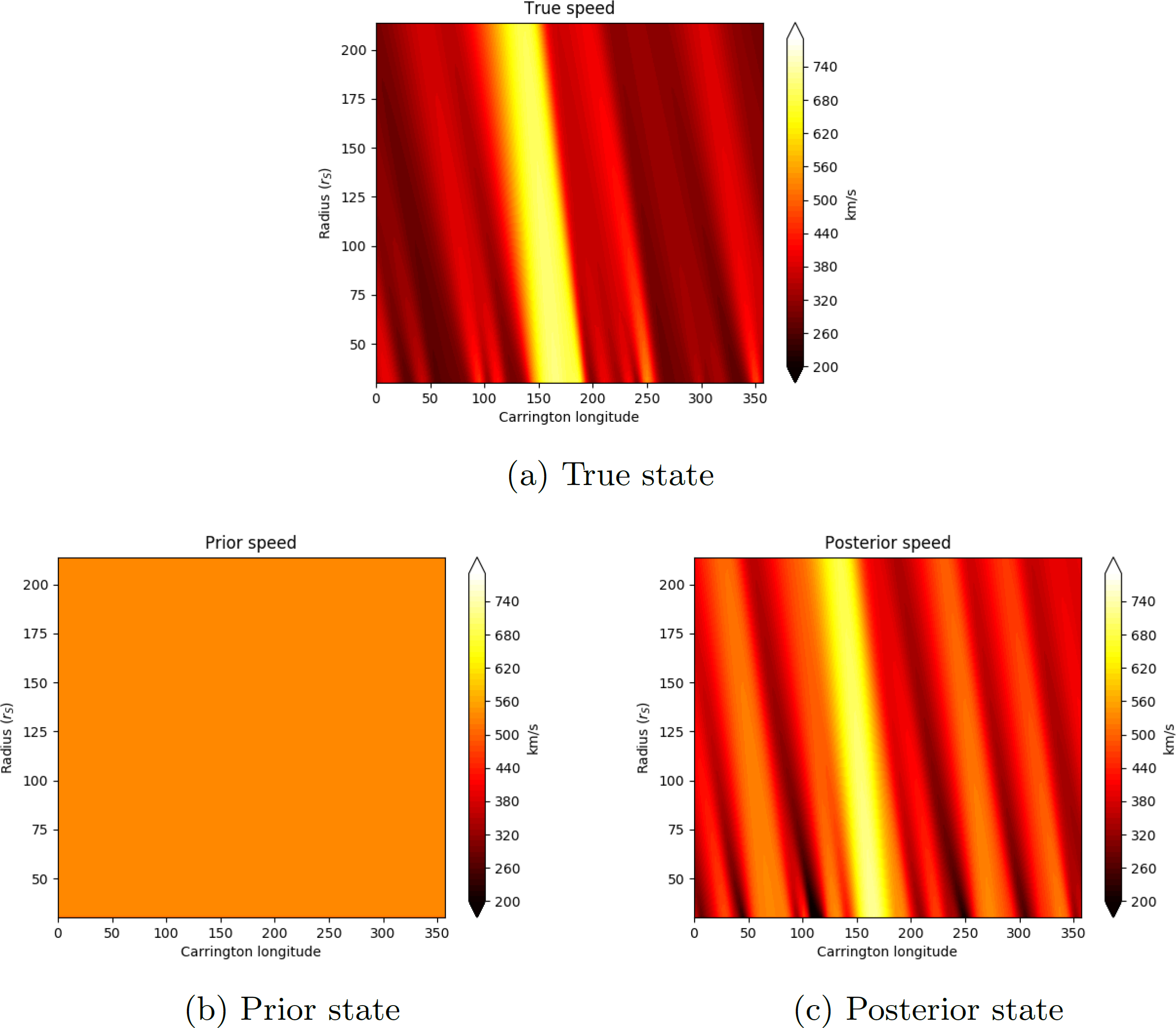}
			\caption{Solar wind speed in $km/s$ propagated from the $30r_S$ to the $215r_S$ using the solar wind propagation model. $(a)$ solar wind speeds generated from the truth state, $(b)$  the uniform prior state of $500 km/s$ and $(c)$ the posterior state after the data assimilation has been performed.}
	\label{fig:twinPriorStateUniform}
\end{center}
\end{figure}

\begin{figure}
\begin{center}
	\includegraphics[width=\textwidth]{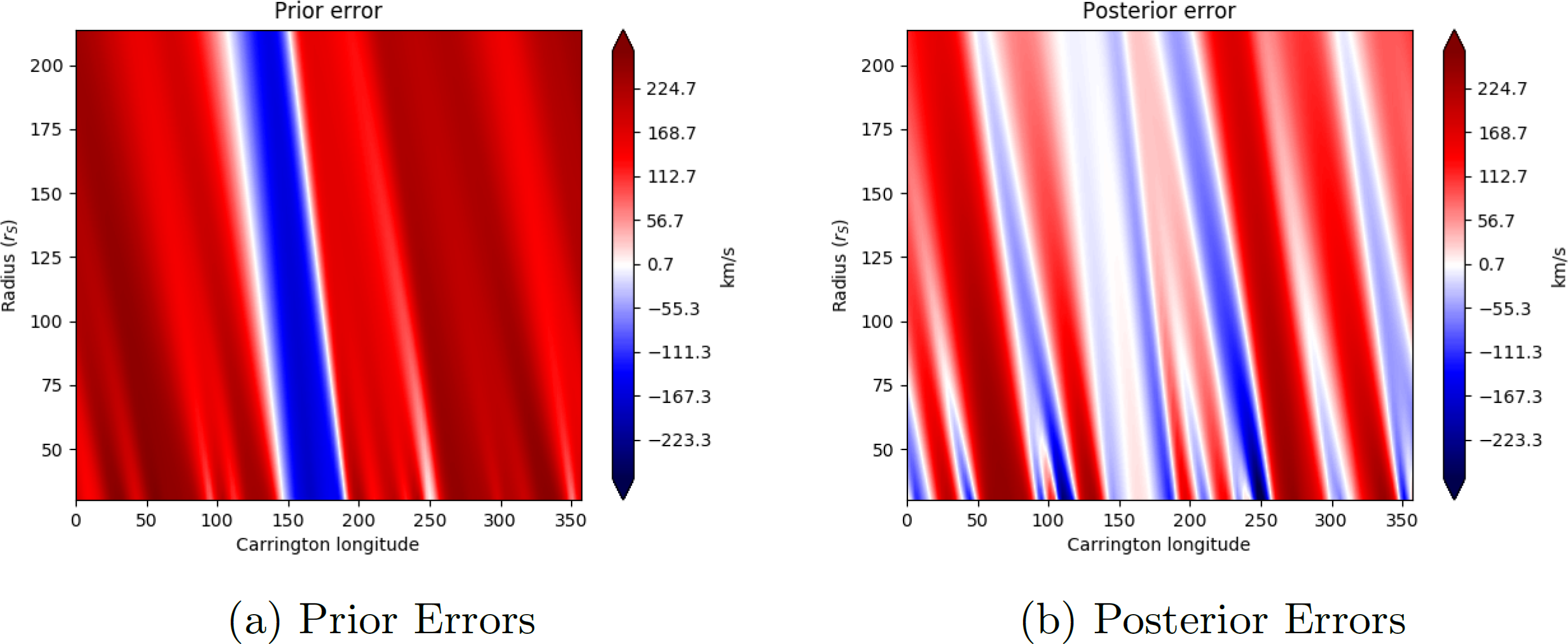}
	\caption{(a)The differences between the prior and the `true' solar wind speeds and (b) the differences between the posterior and the true solar wind speeds, in $km/s$, when the prior state is specified as a uniform speed of $500 km/s$.}
	\label{fig:twinErrorsUniform}
\end{center}
\end{figure}

%
%
%
%

In this section, we consider the effects of different prior states when using data assimilation. These prior states will be generated such that they are no longer random perturbations of the true state (as is commonly the case in all applications, especially as we rarely have a true state to compare to in the first place). This will cause the data assimilation to perform sub-optimally as the assumption regarding the prior state will no longer be valid. Hence the purpose of this section is to show that whilst the data assimilation may be compromised by poorly specified priors, it can still provide significant improvements in the estimation of the true state.

In this section, the true state, prior error covariance matrix, observations and observation error covariance matrix are all specified in the same way as the previous section.

The first ``poor" prior state we use is the prior state generated in the previous section, but shifted arbitrarily. We use an angular shift of $62 \Delta \phi \approx 174^{\circ}$ around the Sun, as this results in a very large difference between the prior and true states. The prior is shown in Figure \ref{fig:twinPriorState62dPhi}b. The fast wind band is now clearly in the incorrect position in the prior state. This implies that there are far larger errors in our prior state, reflected by the much higher prior cost functions and prior RMSE values in Table \ref{tab:0dPhiPriorTwin} when compared to the unshifted case in the previous section.

After the variational data assimilation method is performed, the RMSEs over the whole domain have been greatly reduced (by approximately $50\%$). The posterior state obtained from the variational data assimilation scheme is shown in Figure \ref{fig:twinPriorState62dPhi}c and detailed in Table \ref{tab:0dPhiPriorTwin}. We can see that a new fast wind band (between $\approx 145^{\circ}-190^{\circ}$) has been included in the posterior state in the correct location, but is slightly narrower than the true state's fast wind band. This results in the reduction of most of the negative prior error in this area. The reason the full band is not generated at the inner boundary may be due to the presence of a small fast wind region centred at $200^{\circ}$ in the prior state, which is still present and enhanced by the variational data assimilation scheme. This small region yielded low prior error at the observation location in near-Earth space, therefore the data assimilation scheme will not have removed it. A more noticeable error in the posterior speed is the remains of the prior's fast wind band, which the data assimilation has not been able to fully remove. This results in a large positive wind bias in the posterior error in this region (although greatly reduced from the initial prior error in this region, as can be clearly seen in Figure \ref{fig:twinErrors62dPhi}). Nevertheless, the near-Earth 'observations' have clearly been able to update the inner boundary conditions for the model and result in a persistent change in the model state, something which was not possible with the ensemble-based Kalman filter approach attempted previously \citep{lang2017SWDA}.

The second "poor" prior state is a constant solar wind speed of $500 km/s$ at all points on the inner boundary (see Figure \ref{fig:twinPriorStateUniform}), which mimics having no near-Sun information about the solar wind speed structures. It can be seen from Figure \ref{fig:twinErrorsUniform} that this estimate contains a large positive bias from the true state at all points, except the fast wind band, which is negatively biased. In this case, the data assimilation still reduces the RMSE over the whole domain (by $\approx 43 \%$, as shown in Table \ref{tab:0dPhiPriorTwin}), but has not been able to fully correct for the large positive biases that were present within the prior error. There are still large positive errors over most regions in the posterior state, albeit reduced compared to the prior state. The posterior state can be seen to recreate the fast wind band in the correct location with low errors present in that region. In addition, some of the slower wind regions present in the true state (at $\approx 40^{\circ}, 130^{\circ}$, $200^{\circ}$ and at $300^{\circ}$) have been recreated, albeit still with large errors present.

The experiments in this section show while the data assimilation is capable of reducing the errors in a prior state, specifying an accurate prior state is nevertheless essential to obtain an optimal analysis. We further note that in the future when such data assimilation is to be used in conjunction with a more costly numerical model, it may not be possible to perform enough iterations for the cost function to converge, and specifying an accurate prior state will be even more vital.

\FloatBarrier

\subsection{Using real in situ observations}

\label{sec:stereo}

\begin{figure}
	\begin{center}
		\includegraphics[width=0.6\textwidth]{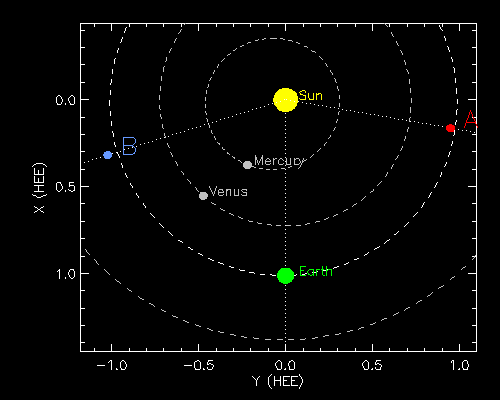}
		\caption{The average STEREO locations for Carrington Rotation 2100, used for the generation of the observation operator, defining the location of the observations.  Figure generated from \textit{https://stereo-ssc.nascom.nasa.gov/where.shtml}}
		\label{fig:stereoLoc}
	\end{center}
\end{figure}

\begin{figure}
	\begin{center}
		\includegraphics[width=0.75\textwidth]{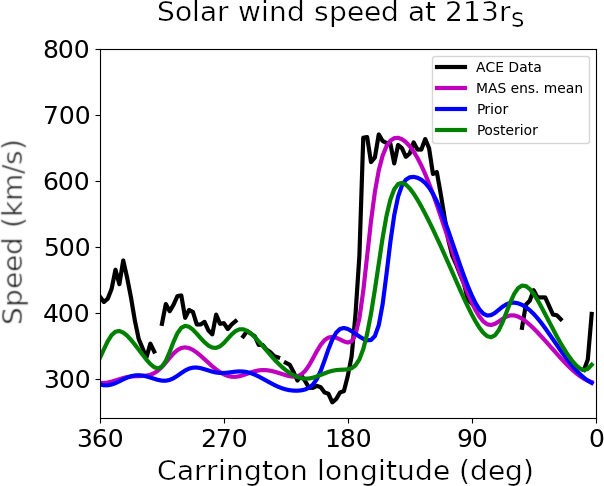}
		\caption{Solar wind speed at $213 r_S$ $(\mathcal{L}1)$ for Carrington Rotation $2100$. The observed solar wind from the ACE spacecraft in near-Earth space is shown in black. The solar wind speeds from the model without any data assimilation (the prior) and after assimilation of STEREO A and B observations (the posterior) are shown in blue and green, respectively. The MAS ensemble mean solar wind speed is shown in magenta. Carrington longitude is shown decreasing to the right, to mimic a time series observed at a fixed point in the heliosphere (e.g., in near-Earth space), assuming the solar wind perfectly co-rotates with the Sun.}
		\label{fig:verifVel}
	\end{center}
\end{figure}

\begin{figure}
	\begin{center}

		\includegraphics[width=\textwidth]{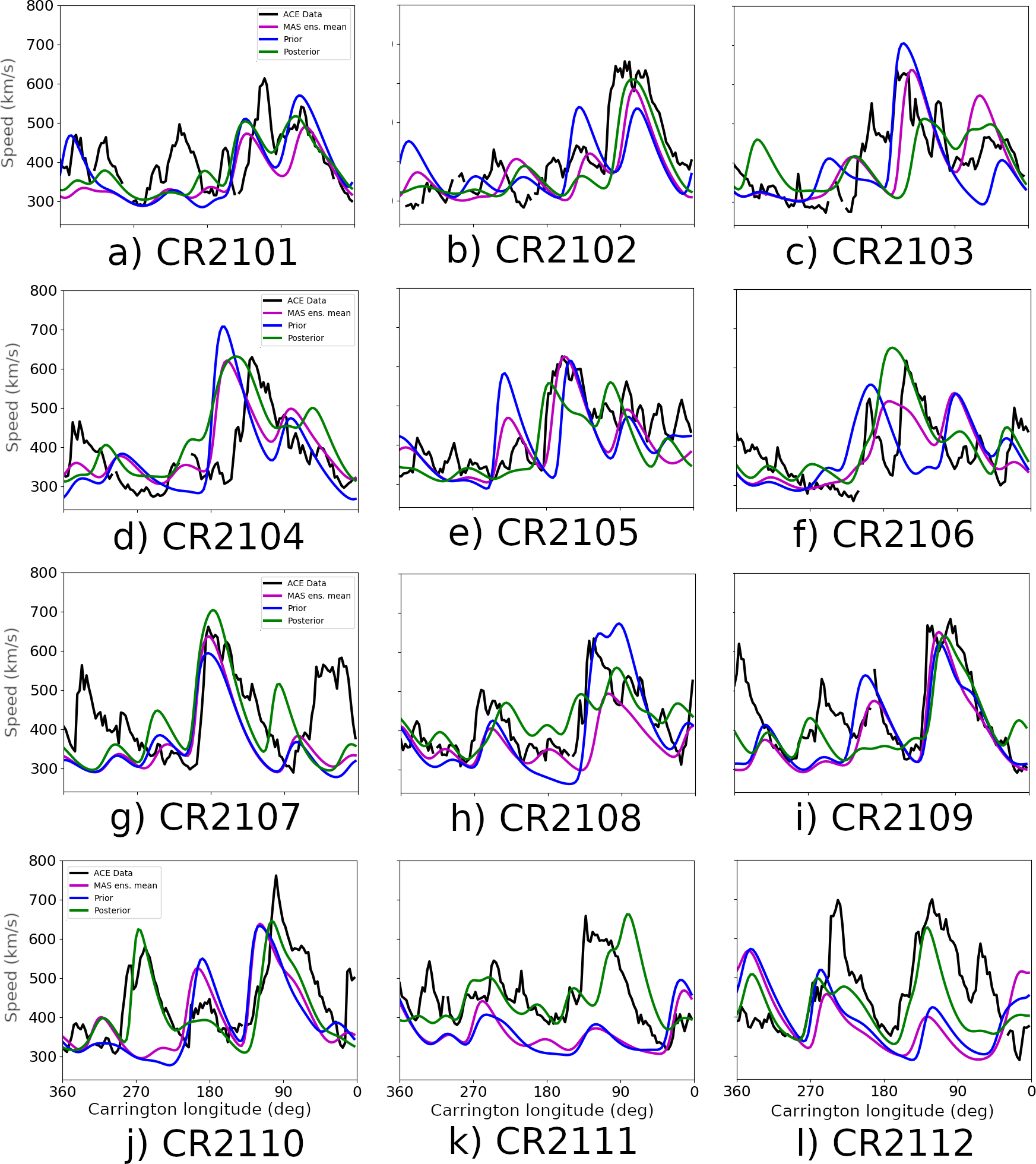}
		\caption{Near-Earth solar wind speed (at $\mathcal{L}1$) for 12 Carrington rotations spanning Oct 2010 to Oct 2011. The black line shows observed solar wind speed in near-Earth space by the ACE spacecraft. The blue lines show the prior state, generated using the MAS ensemble, as specified in \citet{owens2017probabilistic}. The green lines show the posterior state, resulting from assimilation of STEREO A and B observations. The magenta lines show the mean of the MAS ensemble.}
		\label{fig:DA}
	\end{center}
\end{figure}

\begin{center}
	\begin{figure}
		\begin{center}
			\includegraphics[width=0.8\textwidth]{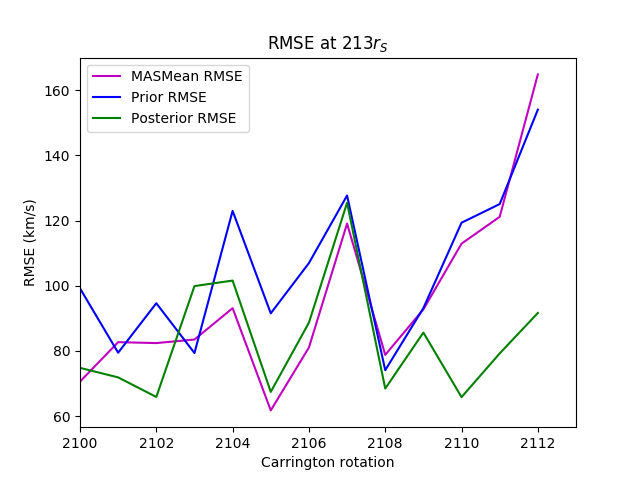}
			\caption{The Root Mean Squared Errors (RMSE) in near-Earth (at $\mathcal{L}1$) solar wind speed as a function of Carrington rotation. Blue line shows the prior state, generated using the MAS ensemble, and the green lines shows the RMSEs of the posterior state, resulting from assimilation of STEREO A and B observations. The magenta line shows the mean of the MAS ensemble.}
			\label{fig:RMSEAll}
		\end{center}
	\end{figure}
\end{center}

\begin{table}
	\begin{center}
		\begin{tabular}{ | p{0.8cm} | p{1.0cm} | p{1.0cm} | p{1.2cm} | p{1.2cm} | p{1.1cm} | p{1.1cm} | p{1.2cm} | p{1.2cm} | p{1.2cm} | }
			\hline
			Carr. Rot & STER. A loc. ($\strut^{\circ}$ from Earth) & STER. B loc. ($\strut^{\circ}$ from Earth) & Init. $\mathcal{J}(\mathbf{v_0})$ & Final $\mathcal{J}(\mathbf{v_0})$ & MAS Mean RMSE (km/s) & Prior RMSE (km/s) & Post. RMSE (km/s) & Post./ MAS Mean red. in RMSE ($\%$) & Post./ Prior red. in RMSE ($\%$) \\ \hline
			2100 & 80.6 & -72.8 & 799.54 & 381.07 & 70.60 & 99.19 & 74.81 & -5.96 & 24.57 \\
			2101 & 82.1 & -76.0 & 918.91 & 570.15 & 82.69 & 79.46 & 71.88 & 13.07 & 9.54\\
			2102 & 83.2 & -79.6 & 1704.32 & 768.73 & 82.41 & 94.64 &  65.89 & 20.05 & 30.38\\
			2103 & 84.1 & -83.6 & 2866.68 & 496.45 & 83.50 & 79.36 & 99.89 & -19.63 & -26.87\\
			2104 & 85.2 & -87.1 & 1435.50  & 444.03 & 93.12 & 122.96 & 101.60 & -9.09 & 17.39\\
			2105 & 86.0 & -90.3 & 2036.20 & 515.61 & 61.76 & 91.54 & 67.44 & -9.20 & 26.32 \\
			2106 & 86.8 & -92.9 & 2429.19 & 598.49 & 81.06 & 106.95 & 88.67 & -9.39 & 17.09\\
			2107 & 87.5 & -94.6 & 1616.64 & 1030.86 & 119.06 & 127.67 & 125.43 & -5.35 & 1.75\\
			2108 & 88.7 & -95.3 & 3086.53 & 1384.29 & 78.78 & 74.09 & 68.47 & 13.09 & 7.59\\
			2109 & 90.6 & -95.0 & 1585.78 & 684.77 & 92.75 & 93.14 & 85.63 & 7.68 & 8.06\\
			2110 & 93.3 & -93.7 & 2756.18 & 740.25 & 112.91 & 119.32 & 65.84 & 41.69 &  44.83\\
			2111 & 96.3 & -92.5 & 4563.84 & 1461.13 & 121.141 & 125.05 & 79.23 & 34.60 & 36.64\\
			2112 & 99.1 & -92.0 & 2353.79 & 1028.48 & 164.82 & 154.00 & 91.65 & 44.39 & 40.49 \\
			\hline
			Avg.  & 87.96 &	-88.11 & 2165.62 &	777.25 & 95.74 & 105.18 & 83.57 & 8.92 & 18.37  \\
			\hline
		\end{tabular}
		\caption{Table showing prior and posterior cost function values, the  RMSEs of the MAS Mean, Prior and Posterior states for each Carrington Rotation and the percentage reductions in RMSE between the MAS Mean/prior state and the posterior states. The bottom row shows the average values across all 13 Carrington rotations.}
		\label{tab:12IterReal}
	\end{center}
\end{table}

In this section, the data assimilation scheme described in Section \ref{sec:DAscheme} is applied to real STEREO \citep{kaiser2008stereo} observations during Carrington Rotation 2100. This Carrington Rotation was chosen because the two STEREO spacecraft have a large separation on either side of the Earth, shown in Figure \ref{fig:stereoLoc}, with STEREO A $80.5^{\circ}$ ahead of the Earth and STEREO B located $72.5 ^{\circ}$ behind. For the purposes of the data assimilation run with the simple solar wind propagation model, the STEREO spacecraft are assumed to be stationary during the Carrington Rotation and located at the same heliographic latitude as Earth. Observations are sampled every $5$ hours,  as this corresponds to the approximate amount of time for the Sun to rotate $\Delta \phi$ degrees and hence matches the implicit time resolution of our solar wind model. We assimilate only the solar wind speed, as that is the only parameter contained within the solar wind model.

The prior error covariance matrix is produced using an ensemble  for Carrington Rotation 2100 via the same methodology as in the OSSE experiments. Similarly, the observation error covariance matrix, $\mathbf{R}$, is generated using equation \eqref{eq:R01xMean}, as in the OSSE experiments.

The prior inner boundary initial condition is drawn from the normal distribution of the MAS ensemble for CR2100, $\mathcal{N}(\mathbf{0},\mathbf{B_{loc}})$ as before (and is not further perturbed in any way). Unlike the OSSE experiments, however, we have no `truth' state to compare, so we test the skill of our data assimilated analysis state against an independent set of observations that are taken from the Advanced Composition Explorer (ACE) spacecraft positioned at $\mathcal{L}1$, approximately $0.99AU \sim 213r_S$ from the Sun on the Earth-Sun line, which are sampled every $5$ hours from OMNI-hourly averages, as for the STEREO A and B spacecraft data.

Figure \ref{fig:verifVel} shows the prior (i.e., model with no data assimilation) and posterior (i.e., model with STEREO A and B observations assimilated) states in comparison to solar wind speed observed at the ACE spacecraft. The band of `fast wind', approximately $600 km s^{-1}$, in the prior state is slower than is observed. But in general, the prior is in relatively good agreement with observations and thus improving on this forecast provides a challenge for the DA scheme. After the DA scheme is used, the posterior state still underestimates the peak speed of the fast wind stream. The greatest improvements achieved by the DA are at the beginning of the Carrington Rotation (between $~315^{\circ}-200^{\circ}$), where the posterior is in much better agreement with the observed ACE data. The smoothness of the prior and posterior states is a reflection of simplicity of the numerical model used to represent the solar wind. The relatively shallow gradient of the fast wind onset in the posterior solution also suggests numerical diffusion may be an issue.

In addition, from the first row of Table \ref{tab:12IterReal}, we see that the overall errors in the system, as denoted by the cost function, have been reduced by $\approx 52 \%$. The RMSE is computed between the model and ACE solar wind speed, such that:
\begin{equation}
\label{eq:RMSEACE}
RMSE=\sqrt{\frac{1}{N_{ACE}}\sum_{i=1}^{N_{ACE}}\left[\left(y_i-x_i\right)^{2}\right]}
\end{equation}
where the $y_i$ and $x_i$ correspond to the ACE observations and state, at their observation time/location.

It can be seen that assimilating the STEREO A and B observations has resulted in a reduction in the near-Earth solar wind speed RMSE of $\approx 25 \%$ over the whole Carrington Rotation. While the prior state is a good representation of the true state in this instance, the variational scheme was still able to produce a lasting change in the model state by correctly updating the inner boundary condition and improving the near-Earth solar wind speed. For this particular Carrington rotation, the MAS ensemble mean (without STEREO data assimilation) provides a very good match to the observed near-Earth solar wind. In particular, the peak of the fast wind band (between $160^{\circ}-90^{\circ}$), is better reproduced by the MAS ensemble mean than by the posterior. The posterior does better capture the structures present in the observed solar wind speed between $360^{\circ}-210^{\circ}$ and $90^{\circ}-0^{\circ}$. However, the overall RMSE of the ensemble mean is lower than for the posterior. Thus, in this instance, the improvements to the prior state provided by the assimilation of STEREO data are not as great as considering the full ensemble, suggesting the prior error covariance matrix for this Carrington rotation is underestimated, perhaps due to over-aggressive localisation.

We now expand the analysis to 12 further Carrington rotations spanning the period October 2010 to October 2011. For each Carrington rotation, the STEREO spacecraft positions are updated (see $2^{nd}$ and $3^{rd}$ columns of table \ref{tab:12IterReal}), as are the $\mathbf{B}$ matrices using a new MAS solar wind speed ensemble as described in Section \ref{sec:expsetup}. The observation error covariance matrices, $\mathbf{R}$, are also updated for each Carrington Rotation, and are dependent upon the prior states, that are generated as a random perturbation (distributed by the normal distribution $\mathcal{N}(\mathbf{0},\mathbf{B})$) from the MAS ensemble mean.

The results are shown in Figures \ref{fig:DA} and \ref{fig:RMSEAll}. The cost function values and RMSEs for the MAS ensemble mean, posterior and prior are listed in Table \ref{tab:12IterReal}. For 12 of the 13 Carrington Rotations, the use of data assimilation leads to an improvement in the estimate of the near-Earth solar wind speed compared to the no-DA model state. The reduction in RMSE is in the range $~2\%-41\%$, hence we are seeing a significant benefit to applying data assimilation in the majority of cases. In particular, we note that for CRs 2105 and 2110, "false alarm" high speed streams have been removed by the DA, whereas for CRs 2110, 2111 and 2112, "missed" high speed streams have been captured, albeit a little later than observed. For CR 2111, the lateness of the fast stream may be explained by the extremely fast CME observed by STEREO A at this time, which may significantly influence the data assimilation. For CRs 2111 and CR 2112, a significant systematic offset has also been removed.

In Carrington Rotation $2103$, however, the data assimilated near-Earth solar wind is worse than the prior state, with the posterior state having a $26\%$ higher RMSE compared to the prior state. There are no interplanetary coronal mass ejections (ICMEs) observed in near-Earth space \citep{cane2003inter} or at the STEREO spacecraft during this period.

In comparison with the MAS ensemble mean (the best estimate of the truth from the MAS ensemble prior to DA), the posterior has greater RMSE for Carrington Rotations $2103$, $2104$, $2105$, $2106$ and $2107$. These RMSE increases are relatively small in magnitude compared to the improvements from DA in other CRs, particularly Carrington rotations 2108 through 2112. The high RMSEs relative to the MAS ensemble mean could be a result of the latitudinal difference in the STEREO and ACE spacecraft (which the 2-dimensional model assumes all lie in the equatorial plane), which could lead to sampling of different solar wind structures. If so, this could be at least partly mitigated by the use of a fully 3-dimensional solar wind model, though heliospheric latitudinal localisation may need to be treated in a different manner to the heliospheric longitudinal localisation considered here. The MAS ensemble mean seems to perform better when there are no transients in the solar wind and when the solar wind is relatively steady-state. The data assimilated solar wind speed performs increasingly better than the MAS ensemble mean when there is increased CR to CR variability, as the solar cycle increases towards the end of the period considered.

Averaged across all 13 Carrington rotations considered here, data assimilation of STEREO data results in $18.4\%$ reduction in near-Earth solar wind RMSE compared to the prior state and an $8.9\%$ reduction in RMSE compared to the MAS ensemble mean. In Figures \ref{fig:STERADA} through \ref{fig:STERBRMSEAll} and Tables \ref{tab:A1} and \ref{tab:B2}, we show that at the observation locations themselves, the gains from DA are greater still. For STEREO A there is a $42.7\%$ reduction in the prior RMSE and a $35.0\%$ reduction in the MAS mean RMSE as a result of the data assimilation. For STEREO B, there is a $38.1\%$ reduction in the prior RMSE and a $29.8\%$ reduction in the MAS mean RMSE as a result of the data assimilation.
	
\section{Limitations to the variational approach}

Whilst this study shows there is a great deal of potential available in the use of variational data assimilation when applied to the solar wind, there are many hurdles that still need to be overcome. It is also noted that the adjoint method is an extremely powerful method once generated, though this is a substantially simpler task for the solar wind propagation tool used here than for a full MHD model of the solar wind, as discussed below.

Variational data assimilation methods require the generation of an accurate $\mathbf{B}$ matrix, which may be the reason for the lesser improvements noted in $CR2107-2109$. The prior covariance matrix, $\mathbf{B}$, may not fully represent the errors present in the prior state due to the linear dependence of the ensemble members. This may lead to a prior covariance matrix that is of too low rank for the gradient to be calculated accurately by the adjoint method. As an alternative to using a different covariance matrix for each Carrington Rotation, a `climatological' covariance matrix could be generated using MAS ensembles from multiple Carrington Rotations. This will increase the amount of information contained within the $\mathbf{B}$-matrix and will negate the need for localisation. However, this solution negates the advantage of having a flow-dependent covariance matrix that is specific to each Carrington Rotation. It may be the case that this `flow dependency' is necessary for effective $\mathbf{B}$ matrix generation and that a Carrington-Rotation-specific prior error uncertainty matrix may need to be created for optimal assimilation. This mirrors the `hybrid' data assimilation schemes currently being investigated in numerical weather prediction \citep{chump2015, bonavita2014hybrid}. In both cases, the accuracy of the $\mathbf{B}$-matrix relies both on the accuracy of the coronal model for defining the near-Sun solar wind, and the method of sampling the near-Sun solar wind to produce the ensemble. Further research is required to quantify the potential improvements from using Carrington-rotation specific $\mathbf{B}$ matrices, `climatological' $\mathbf{B}$ matrices and the sensitivity to the accuracy of the coronal model used.

The adjoint model is unique to each numerical model and can be extremely difficult to compute efficiently for high dimensional models. Furthermore, as mentioned in Section \ref{sec:DAscheme}, the variational data assimilation methods rely upon the linearisation of the numerical model. The simple numerical model in this study is reasonably linear, and therefore long assimilation windows of one full Carrington Rotation (27 days) are possible. For more complex, higher dimensional models that are more nonlinear, the assimilation window will need to be much shorter to avoid computational issues. For numerical weather prediction, the typical window length is $6-12$ hours, but it is unclear how long the assimilation window can and should be for complex MHD models, such as Enlil (e.g. \citep{odstrcil2003modeling,odstrcil1999distortion, odstrcil2004numerical}) or EUHFORIA \citep{poedts2017euhforia}. This depends on the spatial resolution of the model, as higher resolutions typically lead to stronger nonlinearities with shorter error growth timescales, and also how close the model is to the true solution. Numerical weather prediction owes much of its forecast accuracy to the enormous amount of observations in each $6$-hour window (approximately $10^7$ observations). This also means that the initial boundary condition is accurate as well, such that the linearisations work well for longer. Unfortunately, the sparsity of observations for space weather means it cannot readily adopt this solution.

\section{Discussion and Conclusions}

The experiments shown in this study are the first, to the authors' knowledge, application of a variational data assimilation (DA) method applied to the solar wind using in-situ spacecraft observations. This DA method maps observational information from $215 r_S$ back towards the Sun, to the inner boundary of the solar wind model at $30 r_S$. Twin experiments showed that this variational DA is able to reduce the error in the dynamical model using 5-hourly observations in the near-Earth space. By improving the inner boundary conditions of the solar wind speed, it is possible to retrieve the structure of the solar wind speed produced by the numerical model in the entire domain between the Sun and Earth (at least within the spacecraft orbital plane). This allows solar wind structures, such as bands of fast and slow solar wind, to be reconstructed by the data assimilation method. 

This variational DA scheme requires an estimate of the prior error uncertainty matrix, $\mathbf{B}$. In meteorological applications, there are many decades of forecasts and reanalysis datasets that enable a more accurate representation of the prior error uncertainty matrix. In this study, we have used an ensemble of inner boundary conditions to estimate the $\mathbf{B}$ matrix, over a single Carrington Rotation. It may be more useful to generate a $\mathbf{B}$ matrix with more ensemble members, over multiple Carrington rotations, to incorporate more information about possible error structures in the solar wind. This has the benefit of incorporating additional information from the coronal model solutions, which generally capture the overall solar wind structure, but suffer from relatively small positional errors. Further research is also required in order to better determine how to accurately generate the observation error covariance matrix, $\mathbf{R}$, and how to properly incorporate representivity errors into this matrix.

By applying the same data assimilation to three different prior states we have demonstrated the importance of a ``good" prior solution from the coronal models. Thus forecast skill gained from assimilation of solar wind observations will further increase with developments in coronal modelling and initialisation, such as from improved photospheric magnetic field characterisation \citep{arge2010air}. In this study, we focussed on using synthetic and real in-situ spacecraft observations, though in principle the same data assimilation framework can be applied to remote measurements of the solar wind, such as interplanetary scintillation and white-light heliospheric imager observations. The difficulty, however, lies in accurately representing the observational error, both in the solar wind speed measurement and in the generation of the observation operator, $\mathcal{H}$ (as both techniques involve some degree of line-of-sight integration). The latter uncertainty could possibly be approximated within a data assimilation approach by weak-but-widespread localisation, but such techniques will form part of a future study.

The DA scheme was also used to make hindcasts of near-Earth solar wind speed by assimilating approximately 1 year of in-situ observations from the STEREO spacecraft, when they were approximately 80$^\circ$ahead and behind the Earth in its orbit with respect to the Sun. The prior state has been generated from the MAS ensemble, thus the information contained within the STEREO A and B observations is not carried on to subsequent Carrington Rotations. The DA analysis can, however, be used to forecast the inner boundary for the subsequent Carrington Rotation. This means that there are two possible estimates of the inner boundary for the subsequent Carrington Rotation, one using information from within the inner boundary (from the MAS model) and one estimate using information at heliocentric distances beyond the inner boundary (from the assimilation of in situ solar wind data). How we consolidate these disparate estimates is a question for future research. Nevertheless, even with the approach used here, the root mean-square error in the near-Earth solar wind speed hindcasts was reduced by 18\% by the use of STEREO data assimilation. 

The interval considered in this study (August 2010 to August 2011) is close to solar minimum, thus the coronal mass ejection rate is much reduced compared to solar maximum. This interval, however, still features 9 interplanetary coronal mass ejections (ICMEs) observed in near-Earth space \citep[][; though none with speeds above $600 km/s$]{cane2003inter}. Given the large angular separation of the STEREO spacecraft from Earth, it is unlikely these same ICMEs would be present in the assimilated data. The localisation required for such transient structures will be significantly different (in both space and time) than for the ambient, steady-state solar wind. For example, assuming the fast solar wind from an ICME persists at the same Carrington longitude for a whole solar rotation will significantly degrade a forecast. Fast ICMEs could result in "missed" high speed streams in the posterior solution. Conversely, ICMEs seen by the STEREO spacecraft which didn't encounter Earth, could result in "false" high speed streams in the posterior solution. With very limited spatial sampling from in situ spacecraft, such transient events of limited spatial extent will be problematic for any data assimilation scheme. The solution to dealing with transient structures will likely be greater observational sampling (both in space and time), such as with the remotely sensed observations discussed above. But weighting the relative merits of the two data types, namely precision point measurements and non-localised synoptic measurements, will require careful attention in the observation error matrix. In practice, however, the importance of in-situ observations, particularly in terms of the magnetic field observations, may mean automated ICME detection algorithms also need to be applied in real time.

The major outstanding issue with using a variational data assimilation scheme in an operational setting is that the adjoint method is not scalable to higher-dimensional or more complex models. The tangent linear and adjoint models are unique to each numerical model and can be extremely difficult and require many human-years to create an efficient scheme (however, once created, the adjoint method is an extremely powerful and efficient tool, as shown in this paper). This means that without substantial investment, it will not be possible to utilise an adjoint model in a full MHD model, such as Enlil or EUHFORIA. This issue indicates that it is perhaps more useful to use the adjoint-based data assimilation methods with smaller, simpler solar wind/MHD models. This approach could be define an optimum set of boundary condition using solar wind observations which can then be used to drive more complex models. For DA within more complex MHD models, in order to map observational information from near-Earth space to the inner-boundary, it is perhaps more informative/useful to use a hybrid data assimilation method, such as Ensemble-4DVar \citep{chumps2017Part1,chumps2017Part2}, or a smoother-based data assimilation method, such as the Iterative Ensemble Kalman Smoother \citep{bocquet2014IEnKS}. These will be tested in future studies.

\appendix
\section{Deriving Strong Constraint Cost Function}
\label{sec:appCostFunc}
In order to combine the observational information within $\mathbf{y}$ with the state vector, $\mathbf{x}$, we appeal to Bayes' Theorem \citep{bayesTheorem1763}, which states that:
\begin{equation}
\label{eq:bayesThm}
p\left(\mathbf{x}|\mathbf{y}\right)=\frac{p\left(\mathbf{y}|\mathbf{x}\right)p\left(\mathbf{x}\right)}{p\left(\mathbf{y}\right)}
\end{equation}
where $p\left(\mathbf{x}|\mathbf{y}\right)$ is the posterior probability distribution function, the probability of the state, $\mathbf{x}$, occurring given the observation, $\mathbf{y}$ and is the distribution we wish to estimate in data assimilation; $p(\mathbf{y}|\mathbf{x})$ is the likelihood distribution, which is the probability of the observation occurring for any given state; $p\left(\mathbf{x}\right)$ is the prior distribution, the probability of the state occurring and $p\left(\mathbf{y}\right)$ is the probability of the observation occurring and is a constant, normalising factor for any given observation.

The posterior distribution gives the full picture of the probability of the state space, given the observational data and is what we wish to estimate. Typically, however, the state dimension is extremely large, for example, for numerical weather prediction the state vector is of the order $10^9$-dimensional vector. This makes it impossible for the full posterior distribution to be computed explicitly. Therefore, assumptions about the prior distributions, the likelihood distributions and the numerical model must be made to simplify the problem. Furthermore, we must define precisely what we mean by the `optimal estimate'.

In the variational framework, `optimal state' is defined as the state which maximises the posterior probability distribution (the mode of the posterior distribution) \citep{telegrand1986}. In addition, the prior, observation and model errors are all assumed to be unbiased Gaussian distributions. This means that the probability distributions of the prior error, $\boldsymbol{\xi_0}$, the observation error, $\boldsymbol{\epsilon_k}$, and the model error, $\boldsymbol{\eta_i}$ probability distributions can be fully defined by their mean and respective error covariance matrices, such that:
\begin{align}
\label{eq:ch2:normalDefnXi}
p(\boldsymbol{\xi_0})&=\left(\sqrt{2 \pi |\mathbf{B}|}\right)^{-N_x}\exp \left(-\frac{1}{2}\boldsymbol{\xi_0}^T \mathbf{B}^{-1}\boldsymbol{\xi_0}\right)\\
\label{eq:ch2:normalDefnEta}
p(\boldsymbol{\eta_i})&=\left(\sqrt{2 \pi |\mathbf{Q_i}|}\right)^{-N_x}\exp \left(-\frac{1}{2}\boldsymbol{\eta_i}^T \mathbf{Q_i}^{-1}\boldsymbol{\eta_i}\right)\\
\label{eq:ch2:normalDefnEpsilon}
p(\boldsymbol{\epsilon_k})&=\left(\sqrt{2 \pi |\mathbf{R_k}|}\right)^{-N_y}\exp \left(-\frac{1}{2}\boldsymbol{\epsilon_k}^T \mathbf{R_k}^{-1}\boldsymbol{\epsilon_k}\right)
\end{align}
where $\mathbf{B}$ is the prior error covariance matrix, $\mathbf{Q_i}$ is the model error covariance matrix for each point $i$ and $\mathbf{R_k}$ is the observation error covariance matrix for the $k^{th}$ observation.

In this study, as we are performing initial data assimilation experiments in the solar wind, we shall make the further assumption that the model error is zero, i.e., the perfect model assumption. This is known as the Strong-Constraint. Whilst it is possible to define the variational problem with model error \citep{lang2016ParsationEst, Scheichl2013}, the so-called Weak-Constraint approach, it is beyond the scope of this paper and will not be discussed here.

As the model error is assumed equal to zero, $\mathbf{x_i}$ can be written explicitly in terms of the initial condition, $\mathbf{x_0}$, such that $\mathbf{x_i}=f_{i-1}\left(f_{i-2}\left(\dots f_{0}\left(\mathbf{x_0}\right)\dots\right)\right)$. Using the probability distribution functions described by equations \eqref{eq:ch2:normalDefnXi}-\eqref{eq:ch2:normalDefnEpsilon}, it is possible to write the likelihood and prior distributions as:
\begin{align}
\label{eq:normalLikelihood}
p(\mathbf{y_0},\dots,\mathbf{y_{N_y}}|\mathbf{x_0})&=\prod_{k=0}^{N_y}\left[\left(\sqrt{2 \pi |\mathbf{R_k}|}\right)^{-N_y} \exp \left(-\frac{1}{2}\left(\mathbf{y_k}-\mathcal{H}_k(\mathbf{x_0})\right)\mathbf{R_k}^{-1}\left(\mathbf{y_k}-\mathcal{H}_k(\mathbf{x_0})\right)\right)\right]\\
\label{eq:normalPrior}
p(\mathbf{x_0})&=\left(\sqrt{2 \pi |\mathbf{B}|}\right)^{-N_x} \exp \left(-\frac{1}{2}\left(\left(\mathbf{x_0}-\mathbf{x^b}\right)^T\mathbf{B}^{-1}\left(\mathbf{x_0}-\mathbf{x^b}\right)\right)\right).
\end{align}
where $\mathcal{H}_k$ now implicitly contains the numerical model $f_i$.

As $p(\mathbf{y})$ is constant, this implies that the posterior probability can be written as:
\begin{equation}
\label{eq:postDist}
p(\mathbf{x_0}|\mathbf{y_0},\dots,\mathbf{y_{N_y}}) \propto \exp \left(-\mathcal{J}\left(\mathbf{x_0}\right)\right)
\end{equation}
where
\begin{equation}
\label{eq:costFunc}
\mathcal{J}\left(\mathbf{x_0}\right)=\frac{1}{2}\left(\mathbf{x_0}-\mathbf{x^b}\right)^T\mathbf{B}^{-1}\left(\mathbf{x_0}-\mathbf{x^b}\right)+\frac{1}{2}\sum_{k=0}^{N_y}\left[\left(\mathbf{y_k}-\mathcal{H}_k(\mathbf{x_0})\right)\mathbf{R_k}^{-1}\left(\mathbf{y_k}-\mathcal{H}_k(\mathbf{x_0})\right)\right]
\end{equation}
is the cost function. It can be seen that by minimising $\mathcal{J}\left(\mathbf{x_0}\right)$, we maximise the posterior probability distribution, which is what we wish to find.

There are many minimisation algorithms available to do this. The majority of the efficient methods to do this require the gradient/Hessian to be computed, which are often the most difficult parts to obtain estimates for, especially in higher dimensions.

\newpage
\section{Plots/tables of Solar wind speeds at STEREO A and B, over Carrington Rotations 2101-2112}
\FloatBarrier
\subsection{STEREO A}

\begin{figure}[h!]
	\begin{center}
		\includegraphics[width=\textwidth]{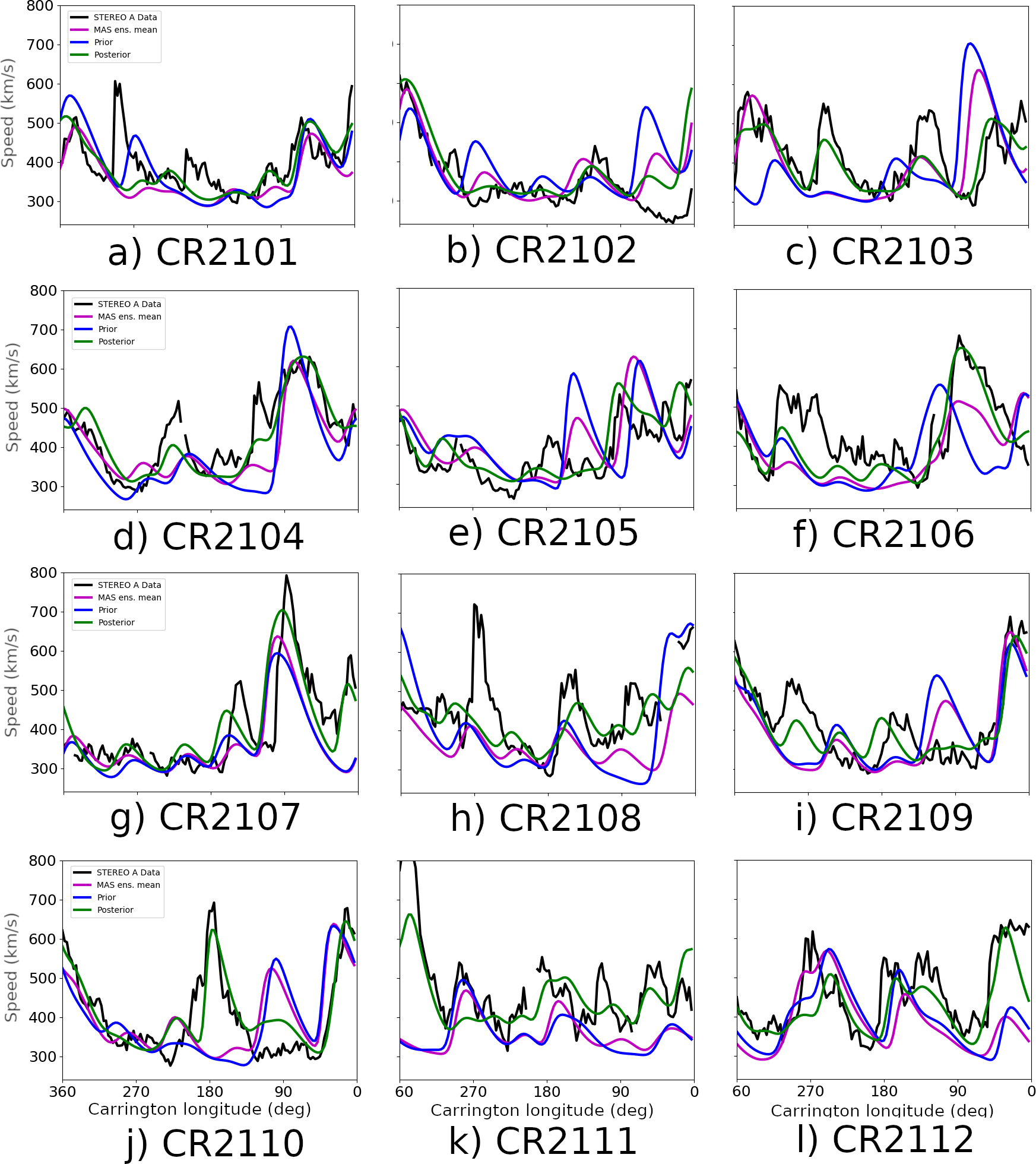}
		\caption{Near-Earth solar wind speed for 12 Carrington rotations spanning Oct 2010 to Oct 2011. The black line shows observed solar wind speed at the STEREO A satellite location. The blue lines show the prior state, generated using the MAS ensemble, as specified in \citet{owens2017probabilistic}, and the magenta line is the MAS ensemble mean. The green lines show the posterior state.}
		\label{fig:STERADA}
	\end{center}
\end{figure}

\begin{center}
	\begin{figure}
		\begin{center}
			\includegraphics[width=0.8\textwidth]{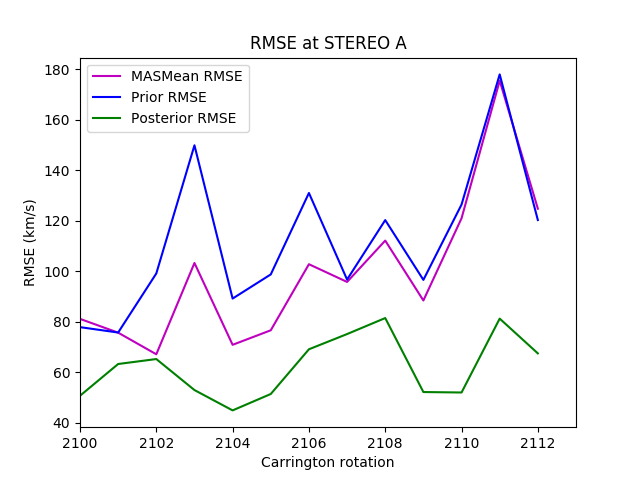}
			\caption{The Root Mean Squared Errors (RMSE) in near-Earth solar wind speed as a function of Carrington rotation. Blue line shows the prior state, generated using the MAS ensemble, and the green lines shows the RMSEs of the posterior state, resulting from assimilation of STEREO A and B observations. }
			\label{fig:STERARMSEAll}
		\end{center}
	\end{figure}
\end{center}

\begin{table}
	\begin{center}
		\begin{tabular}{ | p{0.8cm} | p{1.0cm} | p{1.0cm} | p{1.2cm} | p{1.2cm} | p{1.45cm} | p{1.45cm} | }
			\hline
            Carr. Rot. & STER A pos. & MAS RMSE (km/s) & Prior RMSE (km/s) & Post RMSE (km/s) & MAS/ Prior RMSE Red.($\%$) & Post/ Prior RMSE Red.($\%$)\\ \hline
            2100 & 80.6 & 81.12 & 77.85 & 50.68 & 37.52 & 34.90\\
            2101 & 82.1 & 75.66 & 77.72 & 63.25 & 16.40 & 18.62\\
            2102 & 83.2 & 67.11 & 99.11 & 65.22 & 2.82 & 34.19\\
            2103 & 84.1 & 103.26 & 149.86 & 52.94 & 48.73 & 64.67\\
            2104 & 85.2 & 70.86 & 89.17 & 44.88 & 36.66 & 49.67\\
            2105 & 86 & 76.61 & 98.73 & 51.37 & 32.95 & 47.97\\
            2106 & 86.8	& 102.77 & 130.98 & 69.08 & 32.78 & 47.26\\
            2107 & 87.5	& 95.75 & 96.69 & 75.11 & 21.56 & 22.32\\
            2108 & 88.7	& 112.13 & 120.24 & 81.45 & 27.36 & 32.26\\
            2109 & 90.6	& 88.42 & 95.56 & 52.15 & 41.02 & 45.43\\
            2110 & 93.3	& 120.96 & 126.39 & 51.97 & 57.04 & 58.88\\
            2111 & 96.3	& 175.45 & 177.94 & 81.23 & 53.70 & 54.35\\
            2112 & 99.1	& 124.77 & 120.27 & 67.46 & 45.93 & 43.91\\ \hline
            Avg. & 87.96 & 99.61 & 112.35 & 62.06 & 34.96 & 42.65\\ \hline
		\end{tabular}
        \caption{Table showing the RMSEs of the MAS Mean, Prior and Posterior states for each Carrington Rotation and the percentage reductions in RMSE between the MAS Mean/prior state and the posterior states at STEREO A's locations. The bottom row shows the average values across all 12 Carrington rotations.}
        \label{tab:A1}
    \end{center}
\end{table}

\FloatBarrier
\subsection{STEREO B}

\begin{figure}[h!]
	\begin{center}
		\includegraphics[width=\textwidth]{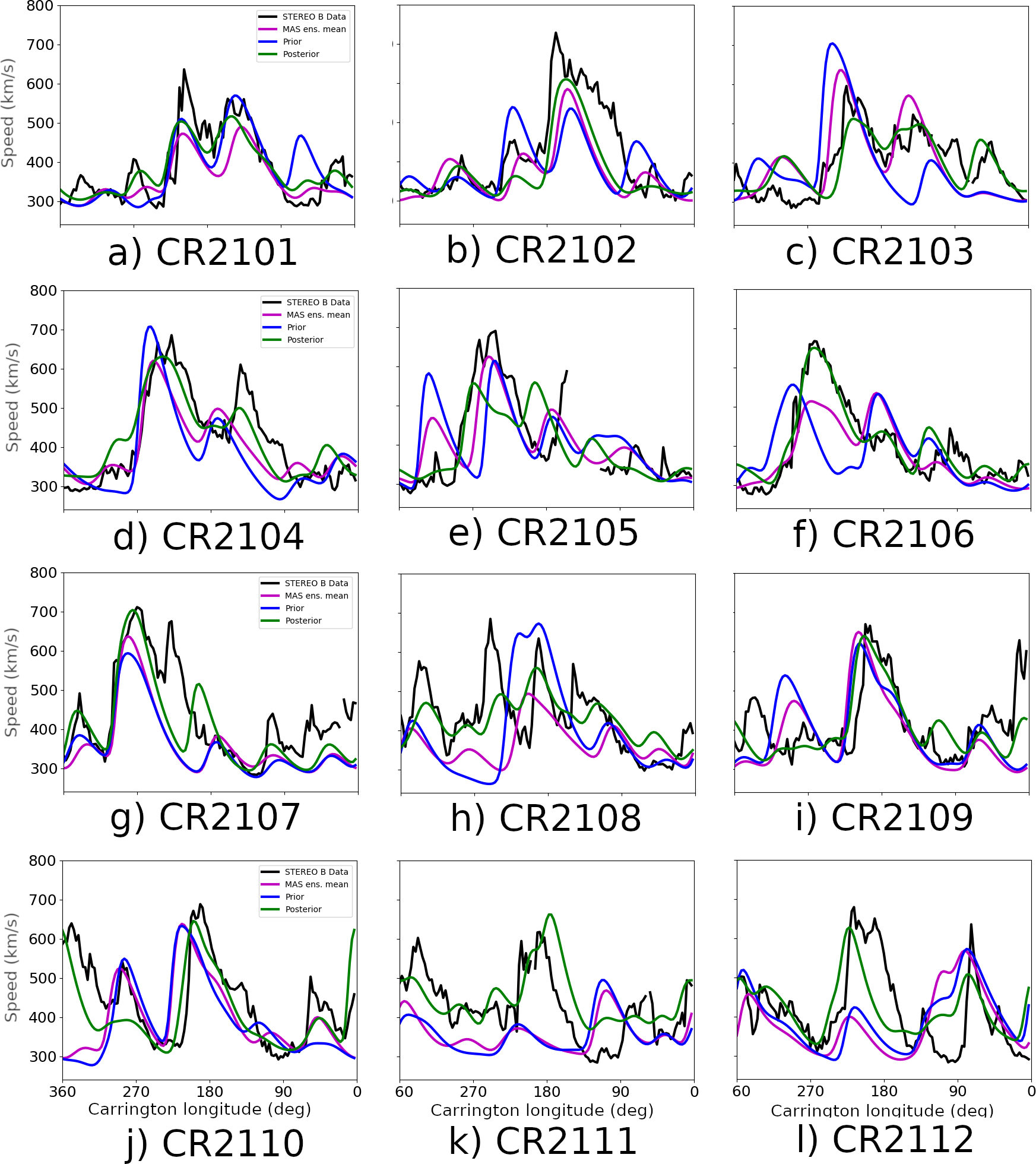}
		\caption{Near-Earth solar wind speed for 12 Carrington rotations spanning Oct 2010 to Oct 2011. The black line shows observed solar wind speed at the STEREO B satellite location. The blue lines show the prior state, generated using the MAS ensemble, as specified in \citet{owens2017probabilistic}, and the magenta line is the MAS ensemble mean. The green lines show the posterior state.}
		\label{fig:STERBDA}
	\end{center}
\end{figure}

\begin{center}
	\begin{figure}
		\begin{center}
			\includegraphics[width=0.8\textwidth]{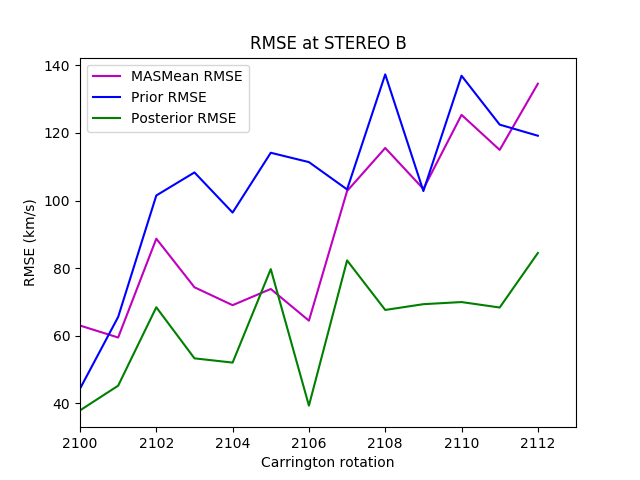}
			\caption{The Root Mean Squared Errors (RMSE) in near-Earth solar wind speed as a function of Carrington rotation. Blue line shows the prior state, generated using the MAS ensemble, and the green lines shows the RMSEs of the posterior state, resulting from assimilation of STEREO A and B observations. }
			\label{fig:STERBRMSEAll}
		\end{center}
	\end{figure}
\end{center}

\begin{table}
	\begin{center}
		\begin{tabular}{ | p{0.8cm} | p{1.0cm} | p{1.0cm} | p{1.2cm} | p{1.2cm} | p{1.45cm} | p{1.45cm} | }
			\hline
			Carr Rot. & STER B pos & MAS RMSE (km/s) & Prior RMSE (km/s) & Post RMSE (km/s) & MAS/ Prior RMSE Red.($\%$) & Post/ Prior RMSE Red.($\%$) \\ \hline
            2100 & -72.8 & 63.02 & 44.34 & 37.92 & 39.83 & 14.48\\
            2101 & -76.0 & 59.48 & 65.56 & 45.23 & 23.96 & 31.01\\
            2102 & -79.6 & 88.71 & 101.49 & 68.43 & 22.86 & 32.57\\
            2103 & -83.6 & 74.33 & 108.31 & 53.32 & 28.27 & 50.77\\
            2104 & -87.1 & 69.04 & 96.43 & 52.06 & 24.59 & 46.01\\
            2105 & -90.3 & 73.83 & 114.12 & 79.71 & -7.96 & 30.15\\
            2106 & -92.9 & 64.47 & 111.36 & 39.33 & 38.99 & 64.68\\
            2107 & -94.6 & 102.80 & 103.28 & 82.28 & 19.96 & 20.33\\
            2108 & -95.3 & 115.54 & 137.30 & 67.63 & 41.47 & 50.74\\
            2109 & -95.0 & 103.43 & 102.80 & 69.33 & 32.97 & 32.56\\
            2110 & -93.7 & 125.33 & 136.90 & 69.97 & 44.17 & 48.89\\
            2111 & -92.5 & 114.96 & 122.44 & 68.35 & 40.54 & 44.18\\
            2112 & -92.0 & 134.54 & 119.16 & 84.47 & 37.22 & 29.11\\ \hline
            Avg. & -88.11 & 91.50 & 104.88 & 62.93 & 29.76 & 38.11\\ \hline
		\end{tabular}
        \caption{Near-Earth solar wind speed for 12 Carrington rotations spanning Oct 2010 to Oct 2011. The black line shows observed solar wind speed at the STEREO B satellite location. The blue lines show the prior state, generated using the MAS ensemble, as specified in \citet{owens2017probabilistic}, and the magenta line is the MAS ensemble mean. The green lines show the posterior state.}
        \label{tab:B2}
    \end{center}
\end{table}
\FloatBarrier
%
%
%
%

%
%
%
%
%
%
%

\acknowledgments
We are grateful to the Space Physics Data Facility and National Space Science Data Center for OMNI. MAS model output is available from the Predictive Science Inc. website: (http://www.predsci.com/mhdweb/home.php). We have benefitted from the availability of Kitt Peak, Wilcox Solar Observatory, Mount Wilson Solar Observatory, SOLIS and GONG magnetograms. MO is part-funded by Science and Technology Facilities Council (STFC) grant number ST/M000885/1.






%
%
%
%
%
%
%
%
%
%

\bibliographystyle{authordate}

\begin{thebibliography}{59}
\providecommand{\natexlab}[1]{#1}
\expandafter\ifx\csname urlstyle\endcsname\relax
  \providecommand{\doi}[1]{doi:\discretionary{}{}{}#1}\else
  \providecommand{\doi}{doi:\discretionary{}{}{}\begingroup
  \urlstyle{rm}\Url}\fi

\bibitem[{\textit{Ades and van Leeuwen}(2013)}]{ades2013exploration}
Ades, M., and P.~J. van Leeuwen (2013), {An exploration of the equivalent
  weights particle filter}, \textit{Quarterly Journal of the Royal
  Meteorological Society}, \textit{139}(672), 820--840, \doi{10.1002/qj.1995}.

\bibitem[{\textit{Amezcua et~al.}(2017)\textit{Amezcua, Goodliff, and {Van
  Leeuwen}}}]{chumps2017Part1}
Amezcua, J., M.~Goodliff, and P.~J. {Van Leeuwen} (2017), {A weak-constraint
  4DEnsembleVar. Part I: formulation and simple model experiments},
  \textit{Tellus A: Dynamic Meteorology and Oceanography}, \textit{69}(1),
  1271,564, \doi{10.1080/16000870.2016.1271564}.

\bibitem[{\textit{Arfken et~al.}(2011)\textit{Arfken, Weber, and
  Harris}}]{arfken2011mathematical}
Arfken, G.~B., H.~J. Weber, and F.~E. Harris (2011), \textit{{Mathematical
  methods for physicists: A comprehensive guide}}, Academic press.

\bibitem[{\textit{Arge et~al.}(2010)\textit{Arge, Henney, Koller, Compeau,
  Young, MacKenzie, Fay, Harvey, Maksimovic, Issautier, Meyer-Vernet,
  Moncuquet, and Pantellini}}]{arge2010air}
Arge, C.~N., C.~J. Henney, J.~Koller, C.~R. Compeau, S.~Young, D.~MacKenzie,
  A.~Fay, J.~W. Harvey, M.~Maksimovic, K.~Issautier, N.~Meyer-Vernet,
  M.~Moncuquet, and F.~Pantellini (2010), {Air Force Data Assimilative
  Photospheric Flux Transport (ADAPT) Model}, in \textit{AIP Conference
  Proceedings}, vol. 1216, pp. 343--346, American Institute of Physics,
  \doi{10.1063/1.3395870}.

\bibitem[{\textit{Bannister}(2007)}]{bannister2007Elementary}
Bannister, R. (2007), {Elementary 4D-Var}, \textit{DARC Technical Report No.2}.

\bibitem[{\textit{Bayes and Price}(1763)}]{bayesTheorem1763}
Bayes, T., and R.~Price (1763), {An Essay towards solving a Problem in the
  Doctrine of Chances. By the late Rev. Mr. Bayes, FRS communicated by Mr.
  Price, in a letter to John Canton, AMFRS}, \textit{Philosophical Transactions
  (1683-1775)}, pp. 370--418.

\bibitem[{\textit{Bazaraa et~al.}(2013)\textit{Bazaraa, Sherali, and
  Shetty}}]{bazarra2013nonlin}
Bazaraa, M.~S., H.~D. Sherali, and C.~M. Shetty (2013), \textit{{Nonlinear
  Programming Theory and Algorithms}}, John Wiley {\&} Sons.

\bibitem[{\textit{Bennett}(1992)}]{bennett1992}
Bennett, A.~F. (1992), \textit{{Inverse Methods in Physical Oceanography}},
  Arnold and Caroline Rose Monograph Series of the American So, Cambridge
  University Press.

\bibitem[{\textit{Bocquet and Sakov}(2014)}]{bocquet2014IEnKS}
Bocquet, M., and P.~Sakov (2014), {An iterative ensemble Kalman smoother},
  \textit{Quarterly Journal of the Royal Meteorological Society},
  \textit{140}(682), 1521--1535, \doi{10.1002/qj.2236}.

\bibitem[{\textit{Bonavita et~al.}(2016)\textit{Bonavita, H{\'{o}}lm, Isaksen,
  and Fisher}}]{bonavita2014hybrid}
Bonavita, M., E.~H{\'{o}}lm, L.~Isaksen, and M.~Fisher (2016), {The evolution
  of the ECMWF hybrid data assimilation system}, \textit{Quarterly Journal of
  the Royal Meteorological Society}, \textit{142}(694), 287--303,
  \doi{10.1002/qj.2652}.

\bibitem[{\textit{Breen et~al.}(2006)\textit{Breen, Fallows, Bisi, Thomasson,
  Jordan, Wannberg, and Jones}}]{breen2006}
Breen, A.~R., R.~A. Fallows, M.~M. Bisi, P.~Thomasson, C.~A. Jordan,
  G.~Wannberg, and R.~A. Jones (2006), {Extremely long baseline interplanetary
  scintillation measurements of solar wind velocity}, \textit{Journal of
  Geophysical Research}, \textit{111}(A8), A08,104, \doi{10.1029/2005JA011485}.

\bibitem[{\textit{Broquet et~al.}(2011)\textit{Broquet, Chevallier, Rayner,
  Aulagnier, Pison, Ramonet, Schmidt, Vermeulen, and
  Ciais}}]{broquet2011european}
Broquet, G., F.~Chevallier, P.~Rayner, C.~Aulagnier, I.~Pison, M.~Ramonet,
  M.~Schmidt, A.~T. Vermeulen, and P.~Ciais (2011), {A European summertime CO 2
  biogenic flux inversion at mesoscale from continuous in situ mixing ratio
  measurements}, \textit{Journal of Geophysical Research: Atmospheres},
  \textit{116}(D23), \doi{10.1029/2011JD016202}.

\bibitem[{\textit{Browne and van Leeuwen}(2015)}]{phil2015HadCM3EWPF}
Browne, P.~A., and P.~J. van Leeuwen (2015), {Twin experiments with the
  equivalent weights particle filter and HadCM3}, \textit{Quarterly Journal of
  the Royal Meteorological Society}, \textit{141}(693), 3399--3414,
  \doi{10.1002/qj.2621}.

\bibitem[{\textit{Browne and Wilson}(2015)}]{browne2015simple}
Browne, P.~A., and S.~Wilson (2015), {A simple method for integrating a complex
  model into an ensemble data assimilation system using MPI},
  \textit{Environmental Modelling {\&} Software}, \textit{68}, 122--128.

\bibitem[{\textit{Bust and Mitchell}(2008)}]{bust2008Ion}
Bust, G.~S., and C.~N. Mitchell (2008), {History, current state, and future
  directions of ionospheric imaging}, \textit{Reviews of Geophysics},
  \textit{46}(1), RG1003, \doi{10.1029/2006RG000212}.

\bibitem[{\textit{Cane and Richardson}(2003)}]{cane2003inter}
Cane, H.~V., and I.~G. Richardson (2003), {Interplanetary coronal mass
  ejections in the near-Earth solar wind during 1996-2002}, \textit{Journal of
  Geophysical Research}, \textit{108}(A4), 1156, \doi{10.1029/2002JA009817}.

\bibitem[{\textit{Clayton et~al.}(2013)\textit{Clayton, Lorenc, and
  Barker}}]{clayton2013operational}
Clayton, A.~M., A.~C. Lorenc, and D.~M. Barker (2013), {Operational
  implementation of a hybrid ensemble/4D-Var global data assimilation system at
  the Met Office}, \textit{Quarterly Journal of the Royal Meteorological
  Society}, \textit{139}(675), 1445--1461, \doi{10.1002/qj.2054}.

\bibitem[{\textit{Courtier et~al.}(1994)\textit{Courtier, Th{\'{e}}paut, and
  Hollingsworth}}]{courtier1994strategy}
Courtier, P., J.~N. Th{\'{e}}paut, and A.~Hollingsworth (1994), {A strategy for
  operational implementation of 4D-Var, using an incremental approach},
  \textit{Quarterly Journal of the Royal Meteorological Society},
  \textit{120}(519), 1367--1387.

\bibitem[{\textit{Dee et~al.}(2011)\textit{Dee, Uppala, Simmons, Berrisford,
  Poli, Kobayashi, Andrae, Balmaseda, Balsamo, Bauer, Bechtold, Beljaars,
  van~de Berg, Bidlot, Bormann, Delsol, Dragani, Fuentes, Geer, Haimberger,
  Healy, Hersbach, H{\'{o}}lm, Isaksen, K{\aa}llberg, K{\"{o}}hler, Matricardi,
  McNally, Monge-Sanz, Morcrette, Park, Peubey, de~Rosnay, Tavolato,
  Th{\'{e}}paut, and Vitart}}]{dee2011eraInterim}
Dee, D.~P., S.~M. Uppala, A.~J. Simmons, P.~Berrisford, P.~Poli, S.~Kobayashi,
  U.~Andrae, M.~A. Balmaseda, G.~Balsamo, P.~Bauer, P.~Bechtold, A.~C.~M.
  Beljaars, L.~van~de Berg, J.~Bidlot, N.~Bormann, C.~Delsol, R.~Dragani,
  M.~Fuentes, A.~J. Geer, L.~Haimberger, S.~B. Healy, H.~Hersbach, E.~V.
  H{\'{o}}lm, L.~Isaksen, P.~K{\aa}llberg, M.~K{\"{o}}hler, M.~Matricardi,
  A.~P. McNally, B.~M. Monge-Sanz, J.-J. Morcrette, B.-K. Park, C.~Peubey,
  P.~de~Rosnay, C.~Tavolato, J.-N. Th{\'{e}}paut, and F.~Vitart (2011), {The
  ERA-Interim reanalysis: configuration and performance of the data
  assimilation system}, \textit{Quarterly Journal of the Royal Meteorological
  Society}, \textit{137}(656), 553--597, \doi{10.1002/qj.828}.

\bibitem[{\textit{{Di Lorenzo} et~al.}(2007)\textit{{Di Lorenzo}, Moore,
  Arango, Cornuelle, Miller, Powell, Chua, and
  Bennett}}]{lorenzo2007weakStrong}
{Di Lorenzo}, E., A.~M. Moore, H.~G. Arango, B.~D. Cornuelle, A.~J. Miller,
  B.~Powell, B.~S. Chua, and A.~F. Bennett (2007), {Weak and strong constraint
  data assimilation in the inverse Regional Ocean Modeling System (ROMS):
  Development and application for a baroclinic coastal upwelling system},
  \textit{Ocean Modelling}, \textit{16}(3-4), 160--187,
  \doi{10.1016/J.OCEMOD.2006.08.002}.

\bibitem[{\textit{Dimet and Talagrand}(1986)}]{telegrand1986}
Dimet, F. X.~L., and O.~Talagrand (1986), {Variational algorithms for analysis
  and assimilation of meteorological observations: theoretical aspects},
  \textit{Tellus A}, \textit{38A}(2), 97--110.

\bibitem[{\textit{Durazo et~al.}(2017)\textit{Durazo, Kostelich, and
  Mahalov}}]{durazo2017local}
Durazo, J.~A., E.~J. Kostelich, and A.~Mahalov (2017), {Local ensemble
  transform Kalman filter for ionospheric data assimilation: Observation
  influence analysis during a geomagnetic storm event}, \textit{Journal of
  Geophysical Research: Space Physics}, \textit{122}(9), 9652--9669,
  \doi{10.1002/2017JA024274}.

\bibitem[{\textit{Errico}(1997)}]{errico1997adjoint}
Errico, R.~M. (1997), {What is an adjoint model?}, \textit{Bulletin of the
  American Meteorological Society}, \textit{78}(11), 2577--2591.

\bibitem[{\textit{Evensen et~al.}(1998)\textit{Evensen, Dee, and
  Schr{\"{o}}ter}}]{parEstEvensen1998}
Evensen, G., D.~P. Dee, and J.~Schr{\"{o}}ter (1998), {Parameter estimation in
  dynamical models}, in \textit{Ocean Modeling and Parameterization}, pp.
  373--398, Springer.

\bibitem[{\textit{Eyles et~al.}(2009)\textit{Eyles, Harrison, Davis, Waltham,
  Shaughnessy, Mapson-Menard, Bewsher, Crothers, Davies, Simnett, Howard,
  Moses, Newmark, Socker, Halain, Defise, Mazy, and Rochus}}]{eyles2009}
Eyles, C.~J., R.~A. Harrison, C.~J. Davis, N.~R. Waltham, B.~M. Shaughnessy,
  H.~C.~A. Mapson-Menard, D.~Bewsher, S.~R. Crothers, J.~A. Davies, G.~M.
  Simnett, R.~A. Howard, J.~D. Moses, J.~S. Newmark, D.~G. Socker, J.-P.
  Halain, J.-M. Defise, E.~Mazy, and P.~Rochus (2009), {The Heliospheric
  Imagers Onboard the STEREO Mission}, \textit{Solar Physics}, \textit{254}(2),
  387--445, \doi{10.1007/s11207-008-9299-0}.

\bibitem[{\textit{Fisher et~al.}(2005)\textit{Fisher, Leutbecher, and
  Kelly}}]{fisher2005equiv}
Fisher, M., M.~Leutbecher, and G.~A. Kelly (2005), {On the equivalence between
  Kalman smoothing and weak-constraint four-dimensional variational data
  assimilation}, \textit{Quarterly Journal of the Royal Meteorological
  Society}, \textit{131}(613), 3235--3246, \doi{10.1256/qj.04.142}.

\bibitem[{\textit{Goodliff et~al.}(2015)\textit{Goodliff, Amezcua, and van
  Leeuwen}}]{chump2015}
Goodliff, M., J.~Amezcua, and P.~J. van Leeuwen (2015), {Comparing hybrid data
  assimilation methods on the Lorenz 1963 model with increasing non-linearity},
  \textit{Tellus A}, \textit{67}(0).

\bibitem[{\textit{Goodliff et~al.}(2017)\textit{Goodliff, Amezcua, and {Van
  Leeuwen}}}]{chumps2017Part2}
Goodliff, M., J.~Amezcua, and P.~J. {Van Leeuwen} (2017), {A weak-constraint
  4DEnsembleVar. Part II: experiments with larger models}, \textit{Tellus A:
  Dynamic Meteorology and Oceanography}, \textit{69}(1), 1271,565,
  \doi{10.1080/16000870.2016.1271565}.

\bibitem[{\textit{Hamill et~al.}(2001)\textit{Hamill, Whitaker, and
  Snyder}}]{hamill2001distance}
Hamill, T.~M., J.~S. Whitaker, and C.~Snyder (2001), {Distance-dependent
  filtering of background error covariance estimates in an ensemble Kalman
  filter}, \textit{Monthly Weather Review}, \textit{129}(11), 2776--2790.

\bibitem[{\textit{Howes et~al.}(2017)\textit{Howes, Fowler, and
  Lawless}}]{howes2017strong}
Howes, K.~E., A.~M. Fowler, and A.~S. Lawless (2017), {Accounting for model
  error in strong-constraint 4D-Var data assimilation}, \textit{Quarterly
  Journal of the Royal Meteorological Society}, \textit{143}(704), 1227--1240,
  \doi{10.1002/qj.2996}.

\bibitem[{\textit{Janji{\'{c}} et~al.}(2017)\textit{Janji{\'{c}}, Bormann,
  Bocquet, Carton, Cohn, Dance, Losa, Nichols, Potthast, Waller, and
  Weston}}]{janjic2017representation}
Janji{\'{c}}, T., N.~Bormann, M.~Bocquet, J.~A. Carton, S.~E. Cohn, S.~L.
  Dance, S.~N. Losa, N.~K. Nichols, R.~Potthast, J.~A. Waller, and P.~Weston
  (2017), {On the representation error in data assimilation}, \textit{Quarterly
  Journal of the Royal Meteorological Society}, \doi{10.1002/qj.3130}.

\bibitem[{\textit{Kaiser et~al.}(2008)\textit{Kaiser, Kucera, Davila, Cyr,
  Guhathakurta, and Christian}}]{kaiser2008stereo}
Kaiser, M.~L., T.~A. Kucera, J.~M. Davila, O.~C.~S. Cyr, M.~Guhathakurta, and
  E.~Christian (2008), {The STEREO mission: An introduction}, in \textit{The
  STEREO Mission}, pp. 5--16, Springer.

\bibitem[{\textit{Kalnay}(2003)}]{kalnay2003atmospheric}
Kalnay, E. (2003), \textit{{Atmospheric modeling, data assimilation and
  predictability}}, Cambridge university press.

\bibitem[{\textit{Lang et~al.}(2016)\textit{Lang, van Leeuwen, and
  Browne}}]{lang2016ParsationEst}
Lang, M., P.~J. van Leeuwen, and P.~A. Browne (2016), {A systematic method of
  parameterisation estimation using data assimilation}, \textit{Tellus A},
  \textit{68}(0).

\bibitem[{\textit{Lang et~al.}(2017)\textit{Lang, Browne, van Leeuwen, and
  Owens}}]{lang2017SWDA}
Lang, M., P.~Browne, P.~J. van Leeuwen, and M.~Owens (2017), {Data Assimilation
  in the Solar Wind: Challenges and First Results}, \textit{Space Weather},
  \textit{15}(11), 1490--1510, \doi{10.1002/2017SW001681}.

\bibitem[{\textit{Linker et~al.}(1999)\textit{Linker, Miki{\'{c}}, Biesecker,
  Forsyth, Gibson, Lazarus, Lecinski, Riley, Szabo, and Thompson}}]{linker1999}
Linker, J.~A., Z.~Miki{\'{c}}, D.~A. Biesecker, R.~J. Forsyth, S.~E. Gibson,
  A.~J. Lazarus, A.~Lecinski, P.~Riley, A.~Szabo, and B.~J. Thompson (1999),
  {Magnetohydrodynamic modeling of the solar corona during Whole Sun Month},
  \textit{Journal of Geophysical Research: Space Physics}, \textit{104}(A5),
  9809--9830, \doi{10.1029/1998JA900159}.

\bibitem[{\textit{Lorenc}(1986)}]{lorenc1986analysis}
Lorenc, A.~C. (1986), {Analysis methods for numerical weather prediction},
  \textit{Quarterly Journal of the Royal Meteorological Society},
  \textit{112}(474), 1177--1194.

\bibitem[{\textit{Macbean et~al.}(2016)\textit{Macbean, Peylin, Chevallier,
  Scholze, and Sch{\"{u}}rmann}}]{macbean2016consistent}
Macbean, N., P.~Peylin, F.~Chevallier, M.~Scholze, and G.~Sch{\"{u}}rmann
  (2016), {Consistent assimilation of multiple data streams in a carbon cycle
  data assimilation system}, \textit{Geosci. Model Dev}, \textit{9},
  3569--3588, \doi{10.5194/gmd-9-3569-2016}.

\bibitem[{\textit{Mackay and Yeates}(2012)}]{mackay2012}
Mackay, D., and A.~Yeates (2012), {The Sun's Global Photospheric and Coronal
  Magnetic Fields: Observations and Models}, \textit{Living Reviews in Solar
  Physics}, \textit{9}(1), 6, \doi{10.12942/lrsp-2012-6}.

\bibitem[{\textit{McGregor et~al.}(2011)\textit{McGregor, Hughes, Arge, Owens,
  and Odstrcil}}]{mcgregor2011}
McGregor, S.~L., W.~J. Hughes, C.~N. Arge, M.~J. Owens, and D.~Odstrcil (2011),
  {The distribution of solar wind speeds during solar minimum: Calibration for
  numerical solar wind modeling constraints on the source of the slow solar
  wind}, \textit{Journal of Geophysical Research: Space Physics},
  \textit{116}(A3), \doi{10.1029/2010JA015881}.

\bibitem[{\textit{Nerger et~al.}(2005)\textit{Nerger, Hiller, and
  Schr{\"{o}}ter}}]{nerger2005pdaf}
Nerger, L., W.~Hiller, and J.~Schr{\"{o}}ter (2005), {PDAF-the parallel data
  assimilation framework: experiences with Kalman filtering}, in \textit{Use of
  high performance computing in meteorology: proceedings of the Eleventh ECMWF
  Workshop on the Use of High Performance Computing in Meteorology, Reading,
  UK, 25-29 October 2004/Eds.: Walter Zwieflhofer; Geoge Mozdzynski, Singapore
  [ua]: World}, pp. 63--83.

\bibitem[{\textit{Odstrcil}(2003)}]{odstrcil2003modeling}
Odstrcil, D. (2003), {Modeling 3-D solar wind structure}, \textit{Advances in
  Space Research}, \textit{32}(4), 497--506.

\bibitem[{\textit{Odstrcil and Pizzo}(1999)}]{odstrcil1999distortion}
Odstrcil, D., and V.~J. Pizzo (1999), {Distortion of the interplanetary
  magnetic field by three-dimensional propagation of coronal mass ejections in
  a structured solar wind}, \textit{Journal of Geophysical Research: Space
  Physics}, \textit{104}(A12), 28,225--28,239.

\bibitem[{\textit{Odstrcil et~al.}(2004)\textit{Odstrcil, Riley, and
  Zhao}}]{odstrcil2004numerical}
Odstrcil, D., P.~Riley, and X.~P. Zhao (2004), {Numerical simulation of the 12
  May 1997 interplanetary CME event}, \textit{Journal of Geophysical Research:
  Space Physics}, \textit{109}(A2).

\bibitem[{\textit{Owens and Forsyth}(2013)}]{owens2013heliospheric}
Owens, M.~J., and R.~J. Forsyth (2013), {The heliospheric magnetic field},
  \textit{Living Reviews in Solar Physics}, \textit{10}(5), 5,
  \doi{10.12942/lrsp-2013-5}.

\bibitem[{\textit{Owens and Riley}(2017)}]{owens2017probabilistic}
Owens, M.~J., and P.~Riley (2017), {Probabilistic Solar Wind Forecasting Using
  Large Ensembles of Near-Sun Conditions With a Simple One-Dimensional
  “Upwind” Scheme}, \textit{Space Weather}, \textit{15}(11), 1461--1474,
  \doi{10.1002/2017SW001679}.

\bibitem[{\textit{Owens et~al.}(2017)\textit{Owens, Lockwood, and
  Riley}}]{owens2017global}
Owens, M.~J., M.~Lockwood, and P.~Riley (2017), {Global solar wind variations
  over the last four centuries}, \textit{Scientific Reports}, \textit{7},
  41,548, \doi{10.1038/srep41548}.

\bibitem[{\textit{Pereira et~al.}(2006)\textit{Pereira, Berre, Pereira, and
  Berre}}]{pereira2006ensemble}
Pereira, M.~B., L.~Berre, M.~B. Pereira, and L.~Berre (2006), {The Use of an
  Ensemble Approach to Study the Background Error Covariances in a Global NWP
  Model}, \textit{Monthly Weather Review}, \textit{134}(9), 2466--2489,
  \doi{10.1175/MWR3189.1}.

\bibitem[{\textit{Poedts and Pomoell}(2017)}]{poedts2017euhforia}
Poedts, S., and J.~Pomoell (2017), {EUHFORIA: a solar wind and CME evolution
  model}, \textit{19th EGU General Assembly, EGU2017, proceedings from the
  conference held 23-28 April, 2017 in Vienna, Austria., p.7396}, \textit{19},
  7396.

\bibitem[{\textit{Riley and Lionello}(2011)}]{riley2011mapping}
Riley, P., and R.~Lionello (2011), {Mapping Solar Wind Streams from the Sun to
  1 AU: A Comparison of Techniques}, \textit{Solar Physics}, \textit{270}(2),
  575--592, \doi{10.1007/s11207-011-9766-x}.

\bibitem[{\textit{Riley et~al.}(2012)\textit{Riley, Linker, Lionello, and
  Mikic}}]{riley2012corotating}
Riley, P., J.~A. Linker, R.~Lionello, and Z.~Mikic (2012), {Corotating
  interaction regions during the recent solar minimum: The power and
  limitations of global MHD modeling}, \textit{Journal of Atmospheric and
  Solar-Terrestrial Physics}, \textit{83}, 1--10,
  \doi{10.1016/J.JASTP.2011.12.013}.

\bibitem[{\textit{Riley et~al.}(2015)\textit{Riley, Linker, and
  Arge}}]{riley2015}
Riley, P., J.~A. Linker, and C.~N. Arge (2015), {On the role played by magnetic
  expansion factor in the prediction of solar wind speed}, \textit{Space
  Weather}, \textit{13}(3), 154--169, \doi{10.1002/2014SW001144}.

\bibitem[{\textit{Sasaki}(1970)}]{sasaki1970numerical}
Sasaki, Y. (1970), {Numerical Variational Analysis formulated under the
  constraints as determined by longwave equations and a Low-pass filter},
  \textit{Monthly Weather Review}, \textit{98}(12), 884--898,
  \doi{10.1175/1520-0493(1970)098<0884:NVAFUT>2.3.CO;2}.

\bibitem[{\textit{Scheichl et~al.}(2013)\textit{Scheichl, Cullen, Freitag, and
  Kindermann}}]{Scheichl2013}
Scheichl, R., M.~Cullen, M.~A. Freitag, and S.~Kindermann (2013),
  \textit{{Large Scale Inverse Problems : Computational Methods and
  Applications in the Earth Sciences.}}, 216 pp., De Gruyter.

\bibitem[{\textit{Stone et~al.}(1998)\textit{Stone, Frandsen, Mewaldt,
  Christian, Margolies, Ormes, and Snow}}]{stone1998advanced}
Stone, E.~C., A.~M. Frandsen, R.~A. Mewaldt, E.~R. Christian, D.~Margolies,
  J.~F. Ormes, and F.~Snow (1998), {The advanced composition explorer}, in
  \textit{The Advanced Composition Explorer Mission}, vol.~86, pp. 1--22,
  Springer, \doi{10.1023/A:1005082526237}.

\bibitem[{\textit{Tr{\'{e}}molet}(2006)}]{tremolet2006accounting}
Tr{\'{e}}molet, Y. (2006), {Accounting for an imperfect model in 4D-Var},
  \textit{Quarterly Journal of the Royal Meteorological Society},
  \textit{132}(621), 2483--2504.

\bibitem[{\textit{van Leeuwen}(2014)}]{PJ2014particle}
van Leeuwen, P.~J. (2014), {Particle filters for the geosciences},
  \textit{Advanced Data Assimilation for Geosciences: Lecture Notes of the Les
  Houches School of Physics: Special Issue, June 2012}, p. 291.

\bibitem[{\textit{Wang et~al.}(2004)\textit{Wang, Hajj, Pi, Rosen, and
  Wilson}}]{wang2004Iono}
Wang, C., G.~Hajj, X.~Pi, I.~G. Rosen, and B.~Wilson (2004), {Development of
  the Global Assimilative Ionospheric Model}, \textit{Radio Science},
  \textit{39}(1), \doi{10.1029/2002RS002854}.

\bibitem[{\textit{Zhu et~al.}(2016)\textit{Zhu, van Leeuwen, and
  Amezcua}}]{zhu2016implicit}
Zhu, M., P.~J. van Leeuwen, and J.~Amezcua (2016), {Implicit Equal-Weights
  Particle Filter}, \textit{Quarterly Journal of the Royal Meteorological
  Society}.

\end{thebibliography}

\end{document}